\documentclass[12pt,man,floatsintext]{apa6}

\usepackage{times}
\usepackage[english]{babel}
\usepackage[utf8x]{inputenc}
\usepackage{amsmath}
\usepackage{amssymb}
\usepackage{graphicx}
\usepackage{subfig}
\usepackage{caption}
\usepackage{booktabs}
\usepackage{apacite}
\usepackage[prependcaption]{todonotes}
\usepackage{soul} % for strikethrough
\usepackage{url}
\usepackage{verbatim}
\usepackage{longtable}

\usepackage{lineno}
\DeclareMathOperator*{\Expect}{\mathbb{E}}

\newcommand{\beginsupplement}{%
        \setcounter{table}{0}
        \renewcommand{\thetable}{S\arabic{table}}%
        \setcounter{figure}{0}
        \renewcommand{\thefigure}{S\arabic{figure}}%
     }

\graphicspath{{./figs/}}

\title{\textbf{Scaling up Psychology via Scientific Regret Minimization: \\ A Case Study in Moral Decisions}}
\shorttitle{Scientific Regret Minimization}
\threeauthors{Mayank Agrawal*}{Joshua C. Peterson}{Thomas L. Griffiths}
\threeaffiliations{Princeton Neuroscience Institute and Department of Psychology, Princeton University\\ $^{\ast}$Corresponding Author: mayank.agrawal@princeton.edu}{Department of Computer Science, Princeton University}{Departments of Psychology and Computer Science, Princeton University}
\note{Author Note: A preliminary version of this work was presented at the 41st Annual Cognitive Science Conference \protect\cite{agrawal2019using}.}

\abstract{Do large datasets provide value to psychologists? Without a systematic methodology for working with such datasets, there is a valid concern that exploratory analyses will produce noise artifacts rather than true effects. In this paper, we offer a way to enable researchers to systematically build models and identify novel phenomena in large datasets. One traditional approach is to analyze the residuals of models---the biggest errors they make in predicting the data---to discover what might be missing from those models. However, once a dataset is sufficiently large, machine learning algorithms approximate the true underlying function better than the data, suggesting instead that the predictions of these data-driven models should be used to guide model-building. We call this approach ``Scientific Regret Minimization'' (SRM) as it focuses on minimizing errors for cases that we know should have been predictable. We demonstrate this methodology on a subset of the Moral Machine dataset, a public collection of roughly forty million moral decisions. Using SRM, we found that incorporating a set of deontological principles that capture dimensions along which groups of agents can vary (\textit{e.g.} sex and age) improves a computational model of human moral judgment. Furthermore, we were able to identify and independently validate three interesting moral phenomena: criminal dehumanization, age of responsibility, and asymmetric notions of responsibility.}

\keywords{moral psychology, machine learning, decision making, scientific regret}

\begin{document}
\maketitle

\section{Introduction}

The standard methodology in psychological research is to identify a real-world behavior, create a laboratory paradigm that can induce that behavior, and then test a variety of hypotheses on a group of participants. This methodology was first pioneered over one hundred years ago and remains the \textit{de facto} approach today. While it enables researchers to dissociate individual variables of interest, it can also lead to over-fixation on a specific paradigm and the small amount of variations it offers in contrast to more broadly sampling the space of experiments relevant to the behavior of interest. As a result, several researchers have started to call for a shift towards mining massive online datasets via crowdsourced experiments \cite{griffiths2015manifesto,jones2016big,goldstone2016discovering,mcabee2017inductive,paxton2017finding, hartshorne2018critical, awad2018moral, schulz2019structured} because the scale offered by the internet enables scientists to quickly evaluate thousands of hypotheses on millions of participants.

\begin{figure}[!htb]
    \centering
    \includegraphics[width=0.95\linewidth]{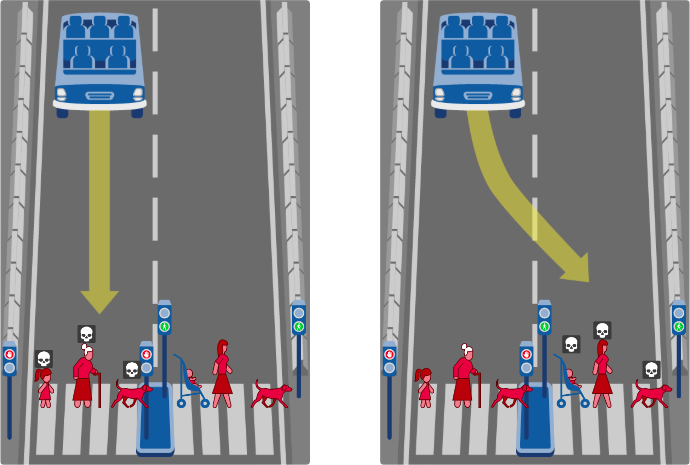}
    \caption{A sample moral dilemma using the Moral Machine paradigm \protect\cite{awad2018moral}. Here, the participant must choose whether an empty car should \textit{stay} and kill a girl, old woman and a dog, who are all illegally crossing, or whether the car should \textit{swerve} and kill an infant, a woman, and a dog, who are all legally crossing.}
    \label{fig:mmsample}
\end{figure}

The Moral Machine experiment \cite{awad2018moral} is one recent example of a large-scale online study. Modeled after the trolley car dilemma \cite{foot1967problem,thomson1984trolley,greene2001fmri}, this paradigm asks participants to indicate how autonomous cars should act when forced to make life-and-death decisions. In particular, participants were presented with two types of dilemmas: pedestrians versus pedestrians, in which an empty car must choose between killing two sets of pedestrians (see Figure \ref{fig:mmsample}), and passengers versus pedestrians (not shown), in which a car must choose between saving its passengers or a group of pedestrians. The Moral Machine experiment collected roughly forty million decisions from individuals in over two hundred countries, making it the largest moral reasoning experiment ever conducted. In addition to the vast number of judgments collected, the experiment operated over a rich problem space: the many possible combinations of twenty different types of agents (\textit{e.g.,} man, girl, female doctor, dog) as well as contextual information (position of the car, crossing signal) resulted in millions of unique dilemmas being presented to participants. With all these variations, the question thus becomes: for any given dilemma, do participants prefer the car to \textit{stay} or \textit{swerve}? Furthermore, what factors influence each decision?

Psychologists have developed a standard statistical approach for analyzing behavioral data to answer such questions: identify all the possible predictors for an individual's decision and fit a model using these predictors. By analyzing the statistical significance of each predictor or an overall model metric that penalizes complexity (\textit{e.g.,} the Akaike information criterion \cite{akaike1998information}), the researcher finds a model that best trades off model complexity with accuracy. Unfortunately, this approach does not scale well with large datasets. Statistical significance is achieved with lower effect sizes in large samples, and complexity penalties are dominated by measures of fit such as the log-likelihood. As a result, when the dataset is sufficiently large, this approach will always favor the more complex model even if the increase in predictive accuracy per data point is trivial, making it difficult to gain insights in the data.

An even stronger critique of this approach is that it assumes prior knowledge of the relevant predictors. In the Moral Machine dataset, the question is not just how important the different factors might be to making moral judgments, but what these factors are to begin with. This suggests the need for exploratory data analysis, a `detective-like' methodology of generating and evaluating hypotheses \cite{tukey1976exploratory,behrens2012exploratory}. One may try to test all possible interactions, but there can easily be an exponential blowup in the number of parameters, reducing the interpretability and thus the explanatory power of the model. For example, a na\"ive featurization of the Moral Machine dataset results in more than $11{,}000$ three-way interactions. Given that the Moral Machine dataset allows forty-way interactions and the relevant predictors may be complex nonlinear functions of the lower-level features, this approach would be difficult to implement in practice. What is needed is an efficient and systematic way of conducting exploratory data analyses in large datasets to identify interesting behaviors and the features that give rise to them.

Understanding the Moral Machine dataset in this manner is simply a microcosm of the broader scientific enterprise. Consider a scientist interested in moral psychology. How does she contribute to the field? She reads papers and combines that knowledge with her own personal experiences, building an internal model that can predict behaviors in different settings. In parallel, she reads the scientific literature to find models that explain these effects. Then, by analyzing the differences between her own mental model and the literature, she either proposes an explanation for a known phenomenon or hypothesizes a novel effect. She conducts an experiment that evaluates her claim and continues this scientific process again.

\begin{figure}[!htb]
    \centering
    \includegraphics[width=\linewidth]{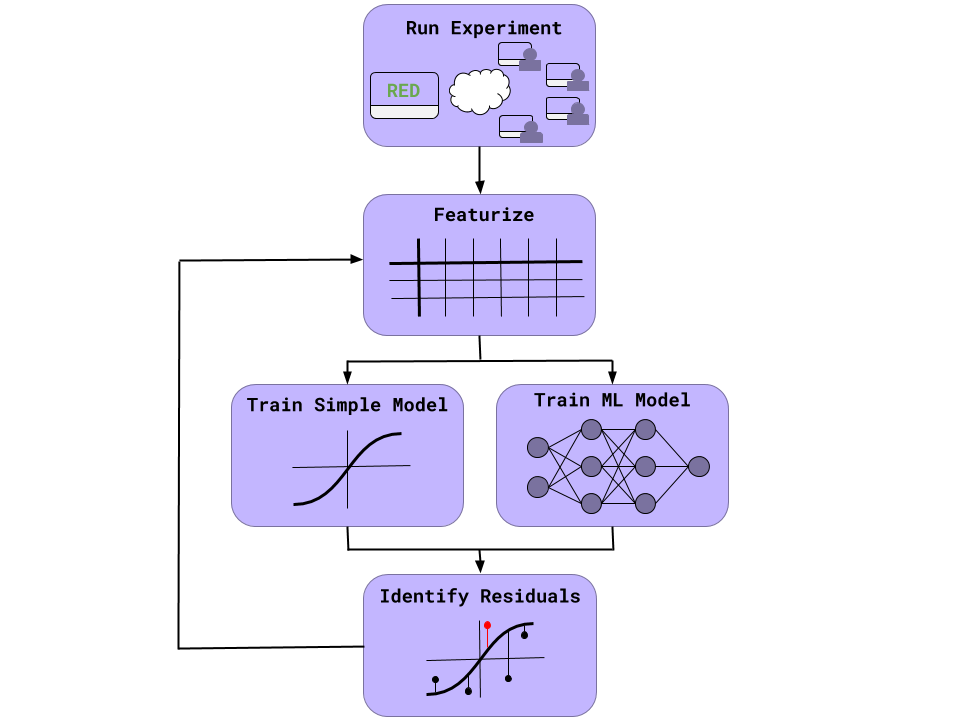}
    \caption{Scientific Regret Minimization. After collecting a large dataset, we use machine learning models to separate the signal from the noise. We then critique psychological models with respect to the signal identified by the machine learning model and continue doing so until both of the models converge.}
    \label{fig:flowchart}
\end{figure}

We believe large datasets should be tackled in the same way, and we formalize this intuition in a process we call ``Scientific Regret Minimization'' (SRM), by analogy to the notion of regret minimization in machine learning \cite{lai1985asymptotically}. First, we suggest that researchers should leverage the size of large datasets to train theoretically-unconstrained machine learning models to identify the amount of variance in the dataset that can be explained \cite{khajah2016deep,peysakhovich2017using,kleinberg2017theory,fudenberg2019predicting,glaser2019roles}. Next, because these models do not necessarily give insight to the underlying cognitive processes, a simple and interpretable psychological model should be fit on the same dataset. Researchers should then critique the psychological model with respect to the black-box model rather than the data. The intuition here is that the psychological model should only be penalized for incorrectly predicting phenomena that are predictable (\textit{i.e.} we should pay close attention to those errors that result in regret). This critiquing process should continue until the predictions of both models converge, thereby ending with a model that jointly maximizes predictive and explanatory power. The residuals from this process may correspond to novel effects, and one can run separate experiments that independently validate them. A summary of this approach is outlined in Figure \ref{fig:flowchart}.

The method of refining models by analyzing their errors (also known as ``residuals'') is often employed in exploratory data analysis \cite{box1962useful,blei2014build,linderman2017using}. In this paradigm, researchers begin by proposing a model and fitting it to the data. By looking at the inputs where the model's predictions and the data diverge, they attempt to identify new relevant features that will hopefully increase the model's accuracy. They then incorporate these new features into the model, fit it to the data, and continue repeating the process. 

Our approach is different because we suggest that, once the dataset is sufficiently large, models should be critiqued with respect to a powerfully predictive \textit{model} rather than the \textit{data}. Critiquing with respect to the data in large datasets can be difficult because the largest residuals often reflect noise. Formally, let $f(x)$ be the true function we are trying to understand, and let the data be $y = f(x) + \epsilon$, where $\epsilon \sim \mathcal{N}(0, \sigma^2_{\epsilon}$). Furthermore, let us assume we are trying to predict the data with a psychological model $g(x)$. The expected squared residual between the psychological model and the data is 
\begin{equation}
    \label{eq:regretderivation}
    \Expect_{p(x, y)}\bigg[y - g(x)\bigg]^2 = \Expect_{p(x)}\bigg[\big(f(x) - g(x)\big)^2\bigg] + \sigma_{\epsilon}^2
\end{equation}
That is, the expected residual between the model and the data, $y - g(x)$, will be the true residual, $f(x) - g(x)$, plus a term that captures the noise variance. (Derivations of all results appear in Materials and Methods). Throughout this paper, we will refer to the residuals between the model and data as the \textit{raw} residuals. Equation \ref{eq:regretderivation} indicates that the correlation between the raw residuals and the true residuals will have an upper bound determined by the noise variance, thus highlighting an important problem with using them to guide model-building. The manual process of critiquing models with respect to the raw residuals often focuses on using the largest $k$ residuals to formulate new predictors. However, as the number of unique inputs increases, these $k$ residuals will mostly reflect noise because $\Expect [\max \vert \epsilon \vert]$ increases as well.

\begin{figure}[!htb]
\centering
\includegraphics[trim={5 0 5 0},clip,width=\textwidth]{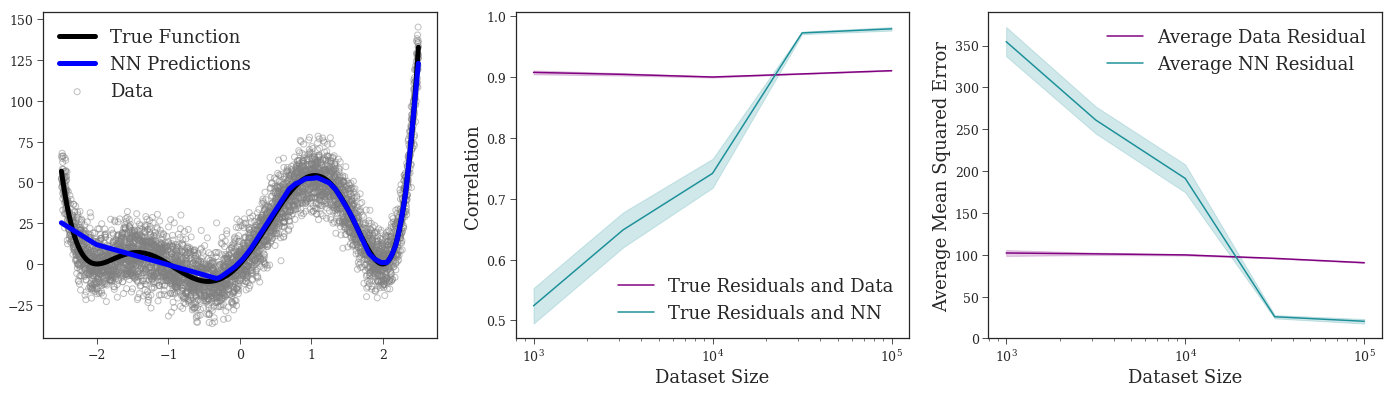}
\caption{Scientific Regret Minimization demonstration. (Left) A graph that outlines the true polynomial function, the data drawn from the polynomial function (with added noise), and a neural network's prediction. (Middle) The correlation between the raw residuals and the true residuals versus the correlation between the smoothed residuals and the true residuals for a simple linear model fit to the data. (Right) The average squared residual between the data and the true function versus the average residual between the neural network and the true function. As predicted, smoothed residuals correlate better with the true residuals when the error of the neural network falls below the noise in the data. Ten simulations were run for each dataset size and error bars in figures the middle and right reflect $\pm1$ SEM.}
\label{fig:regretmin}
\end{figure}

If we think back to our hypothetical scientist, she is analyzing the differences between her internal model and the psychological models in the literature. Once she has read enough of the literature and has enough real-world experience, her internal model will be more sophisticated than a simple table lookup of the data. Formally, let $\widehat{f}(x)$ correspond to a data-driven machine learning algorithm, such as a neural network. The expected residual between this model and the psychological model is
\begin{equation}
\begin{split}
\label{eq:nnregretderivation}
\Expect_{p(x, y)}\bigg[\widehat{f}&(x) - g(x)\bigg]^2 = \Expect_{p(x)}\bigg[\big(f(x) - g(x)\big)^2 + \\ &2\big(f(x) - g(x)\big)\big(\hat{f}(x) - f(x)\big) + \big(\hat{f}(x) - f(x)\big)^2\bigg]
\end{split}
\end{equation}
We will refer to these residuals as the \textit{smoothed} residuals. The latter two terms in the right-hand expression correspond to the covariance of the predictive and psychological models' errors, and the generalization error of the predictive model. When the expression in Equation \ref{eq:nnregretderivation} is less than the expression in Equation \ref{eq:regretderivation}, \textit{i.e.}
\begin{equation}
\Expect_{p(x)}\bigg[2\big(f(x) - g(x)\big)\big(\hat{f}(x) - f(x)\big) + \big(\hat{f}(x) - f(x)\big)^2\bigg] < \sigma_{\epsilon}^2 
\label{residsinequality}
\end{equation}
the smoothed residuals will be more highly correlated with the true residuals than the raw residuals. Because the generalization error of data-driven machine learning algorithms decreases with the amount of the data by which they are trained \cite{huang2018gpipe}, the above inequality will hold when the dataset is sufficiently large. Once this condition is met, we should critique the psychological model with respect to the machine learning model rather than the dataset. Figure \ref{fig:regretmin} demonstrates an example of how smoothed residuals become more representative of the true residuals than the raw residuals as the dataset becomes large. In practice, it is difficult to know when the dataset is large enough for this condition to be reached. For this paper, we approximated it as the point at which the machine learning model outperformed the psychological model.

As a case study, we apply Scientific Regret Minimization to the Moral Machine dataset. We demonstrate that a multilayer feedfoward neural network outperforms simple psychological choice models for predicting people's decisions, and we then continuously critique a rational choice model until its predictive accuracy rivals that of the neural network. The result is an informative, interpretable psychological theory that identifies a set of moral principles that inform people's judgments --  exactly the kind of insight that is relevant to informing policy around new technologies such as autonomous vehicles. This process also allowed us to identify three subtle and complex moral phenomena, which we validated by running preregistered experiments. Our end product is (1) a computational model of moral judgment that jointly maximizes explanatory and predictive power as well as (2) the identification and replication of several principles behind human moral reasoning.

\section{Results}

\subsection{Computational Modeling}

\subsubsection{Formalization}

Scientific Regret Minimization first calls for identifying a paradigm of interest and then critiquing a simple and interpretable psychological model with respect to a data-driven predictive model. We restricted ourselves to the subset of the Moral Machine dataset that contained pedestrians vs. pedestrians dilemmas (N = $15{,}226{,}477$). We used a rational choice model \cite{luce1959individual,mcfadden1973conditional} as our psychological model to explain human moral judgment, assuming that, in the Moral Machine paradigm, humans constructed values for both sides of pedestrians (\textit{i.e.,} $v_{\text{left}}$ and $v_{\text{right}}$) and saved the side with the higher value. Each side's value was determined by aggregating the utilities of its agents:
\begin{equation}
v_{\text{side}} = \sum_iu_il_i
\end{equation}
in which $u_i$ was the utility given to every agent type $i$ (\textit{e.g.,} man, girl, female doctor, dog), and $l_i$ represents the number of those agents on that side. This formalization assumes that a participant's choice $c$ obeys the softmax choice rule, which states that participants chose to save a side in the following way:
\begin{equation}
P(c = \text{left}|v_{\text{left}}, v_{\text{right}}) = \frac{e^{v_{\text{left}}}}{e^{v_{\text{left}}} + e^{ v_{\text{right}}}}
\end{equation}
We implemented this rule by fitting a logistic regression model to the data in order to infer the utility vector $\mathbf{u}$. We called this model the `Utilitarian' model.

This model, however, did not incorporate the main inspiration behind the trolley car dilemma: a resistance to intervening and thus killing bystanders, which is not justified by utilitarian calculus. In order to incorporate such principles, we created a `Deontological' model in which the value of a side is

\begin{equation}
v_{\text{side}} = \sum_m\lambda_mf_m %\sum_iu_il_i + \sum_m\lambda_mf_m 
\end{equation}
Here, $\lambda_m$ refers to the strength of principle $m$ and $f_m$ is a binary variable indicating whether that principle was relevant to the given side. We  proposed that two potential principles were relevant in the Moral Machine paradigm. The first was that a side was penalized if saving it required the participant to swerve. This penalty has been the primary focus of many moral psychology experiments based on the trolley car dilemma \cite{greene2001fmri,greene2004neural,cushman2006role}. Second, because the Moral Machine dataset had three different crossing signal statuses (crossing legally, crossing illegally, and the absence of a crossing signal) we added a penalty if a side's pedestrians were crossing illegally. This side might have been penalized by participants because the participants were waiving their rights to protection by violating the law \cite{nino1983consensual}, and participants may have preferred to kill the pedestrians whose rights have been waived. We used logistic regression to infer the values $\mathbf{\lambda}$.

Lastly, given research demonstrating that individuals have both utilitarian and deontological tendencies \cite{greene2007secret,greene2008cognitive,lombrozo2009role,cushman2013action,crockett2013models}, we created a `Hybrid' model in which the value of a side is a combination of utilitarian and deontological features: 

\begin{equation}
v_{\text{side}} = \sum_iu_il_i + \sum_m\lambda_mf_m 
\end{equation}
This model served as our baseline psychological model for which to iterate upon during SRM.

Central to SRM is that, in addition to training these three choice models, we need to train a data-driven machine learning model. We built a standard multilayer feedforward neural network with forty-two inputs: twenty corresponding to the agents on the left, twenty for the agents on the right, one for the car side, and one for the crossing signal status, thus completely specifying the given dilemma. (It should be noted that one variable for the crossing signal status of the left-hand side is sufficient because the crossing signal status of the right-hand side is just the opposite). These inputs were the same as the `Hybrid' model, except that the `Hybrid' model had the added constraint that the value of an agent was constant across both sides (\textit{i.e.,} a girl on the left side was just as valuable as a girl on the right side), while the neural network had no restriction. 

Finally, as a comparison to a standard data analysis method, we applied a Bayesian variable selection method to a model that started off with all features given to the `Hybrid' model as well as all two- and three-way interactions. Further details about this model are outlined in the SI Text.

% This model had more than $1,500$ parameters, and its formulation and training procedure are described in the Supplementary Information.

\begin{figure}[!htb]
    \centering
    \includegraphics[width=0.32\textwidth]{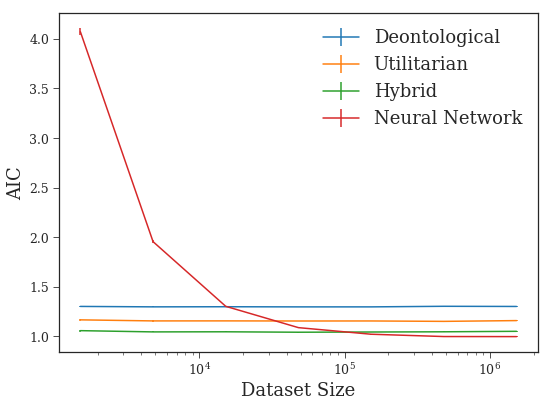}
    \label{ex0}
    \includegraphics[width=0.32\textwidth]{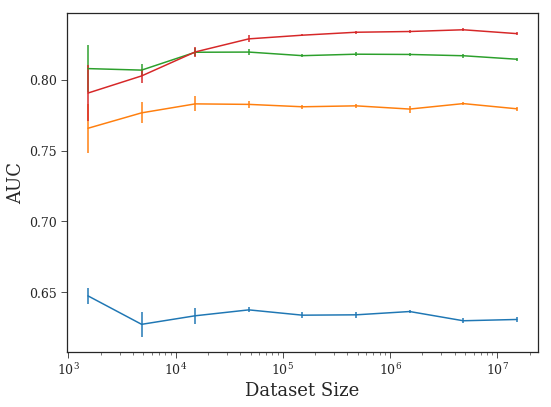}
    \label{ex1}
    \includegraphics[width=0.32\textwidth]{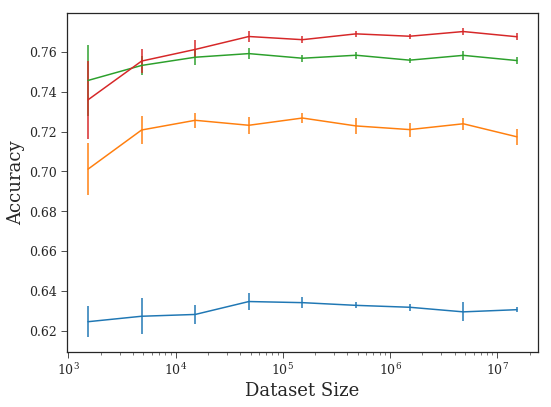}
    \label{ex2}
    \includegraphics[width=0.32\textwidth]{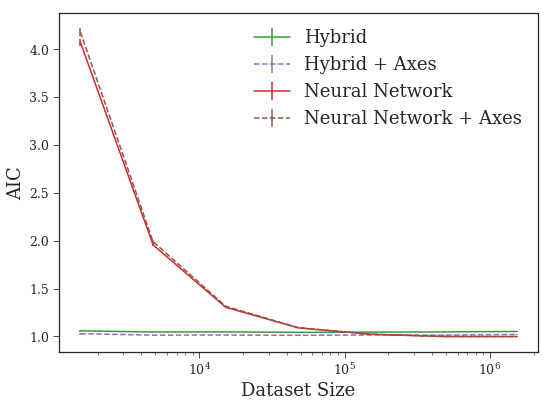}
    \label{ex0_pt}
    \includegraphics[width=0.32\textwidth]{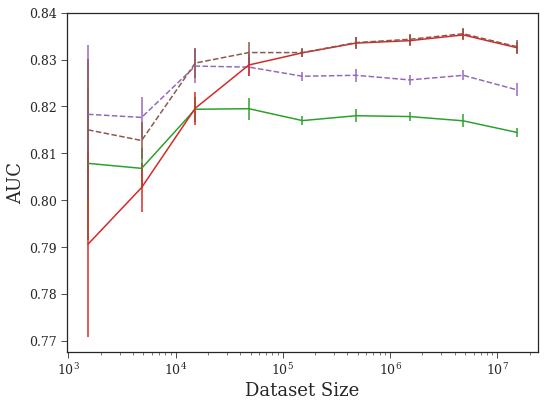}
    \label{ex1_pt}
    \includegraphics[width=0.32\textwidth]{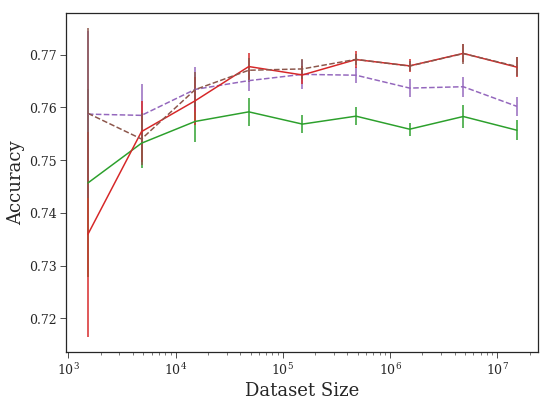}
    \label{ex2_pt}
    \caption{Metrics for Different Models Trained on Subsets of the Moral Machine Data. (Top) Performance of initial choice models and neural network as a function of dataset size. Five bootstrapped samples were taken for every dataset size. Error bars indicate $\pm$1 SEM. (Bottom) Comparison of a choice model and a neural network before incorporating axes of differences versus after incorporating axes of differences. The addition of these features resolves much of the gap between the choice model and the neural network. Error bars indicate $\pm$1 SEM.}
    \label{fig:initmetrics}
\end{figure}

\subsubsection{Initial Results}

The top panel of Figure \ref{fig:initmetrics} reports the results of training all the models on differently-sized subsets of the data. Each model was trained on eighty percent of the subsets, and the metrics here reflect the results when tested on the held-out twenty percent. This procedure was completed for five different splits of the data. We report accuracy and area under the curve (AUC), two commonly used metrics in evaluating models of binary decisions. Furthermore, we also calculated the normalized Akaike information criterion (AIC), a metric in which a smaller number suggests a better model \cite{akaike1998information}.

In this training, the rational choice models performed extremely well at small sizes, and their performance stayed relatively consistent as the dataset size increased. On the other hand, the neural network performed poorly at small sizes, but became better with larger ones and eventually surpassed the choice models\footnote{\ Furthermore, it should be noted that many modern neural networks have problems with calibration even when they have a better AUC  \cite{guo2017calibration}. We thus computed a calibration plot in Supplementary Figure S1 to ensure the neural network served as a good predictive model.}. We also want to point out the Neural Network had a better AIC than the `Hybrid' model despite the fact the former had over three thousand parameters while the latter only had twenty-two. This result affirms our earlier point that metrics like the AIC become uninformative, reducing to a measure of the log-likelihood, when the dataset is sufficiently large.

Most importantly, the neural network's eventual performance suggested there were systematic effects that our choice models were predicting incorrectly. We leveraged these residuals via SRM to build a better choice model of human moral judgment.

\subsection{Improving the Model}

\subsubsection{Identifying Axes of Differences}

The standard methodology for critiquing models suggests prioritizing the \textit{raw} residuals, the largest differences between the choice model and the data. Table \ref{tab:datresid} reports the five largest of these with a minimum sample size of one hundred participants. We claim that the residuals for these dilemmas may often reflect noise and that the neural network's predictions are more representative of the true function than the data. For example, in the largest raw residual, a car is headed towards a group of four humans (a man, a woman, a girl, and a male executive). On the other side is a dog and three cats. According to the data, over $99\%$ of the 649 participants in this dilemma stayed in the lane and chose to kill the humans instead of the animals. The choice model predicted a strong effect in the opposite direction, and this prediction was reasonably close to the neural network's prediction, suggesting that the choice model may not be mispredicting here. To confirm this, we looked at the dilemmas that followed these conditions: the car was headed towards agents that were comprised of men, women, girls, male executives, or any combination of them; the other side comprised of dogs and/or cats; there was an absence of a crossing signal; the number of agents on each side were identical; and at least fifty participants responded to the dilemma. There were forty-five such dilemmas. In forty-four of these forty-five dilemmas, only $11.3\%$ to $25.5\%$ of participants chose to kill the side with humans. The forty-fifth dilemma was the one with the largest residual, and here $99.4\%$ of participants chose to kill the human side. The results of the forty-four other dilemmas suggest that the data for this dilemma is noisy, and thus we shouldn't critique the choice model for disagreeing with the data here.

Similarly, consider the second largest raw residual. Here, a car is headed towards an old woman and a pregnant woman, who are crossing illegally. On the other side is a dog and cat crossing legally. Both the data and the neural network predicted participants would not kill the humans. However, the magnitudes were drastically different, and the correct magnitude is needed to understand the priority of this residual. In the data, only $5.1\%$ of the $924$ participants killed the humans, while the neural network predicted $25.8\%$ of participants would. Like above, we conducted an analysis of the data in similar dilemmas. We looked at dilemmas in which the car was headed towards agents that were either pregnant women, old women, or both; the pedestrians in front of the car were crossing illegally; on the other side of the car were animals; the number of agents on the left and right side were equivalent; and at least fifty people responded to the dilemma. In twelve of the thirteen dilemmas, $14.7\%$ to $35.8\%$ of participants chose to kill the side with humans. The thirteenth was the dilemma reflected here, and thus the data of similar dilemmas suggests the neural network's prediction is more accurate than the data. Therefore, while this dilemma exhibits a large residual for the choice model, the magnitude of the residual is overestimated when critiquing with respect to the data.

\begin{table}[!htb]
    \centering
    \caption{Biggest Differences Between Choice Model and Data (proportions show observed or predicted proportion killing left side).}
    \begin{tabular}{p{0.75in}cccc}%cccc}
    \toprule
    \parbox[c]{1em}{\includegraphics[width=0.75in]{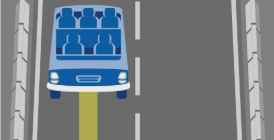}} & \textbf{N} & \textbf{Data}    & \textbf{Choice Model}  & \textbf{Neural Network}  \\[0.2cm]
    \midrule
    \parbox[c]{1em}{\includegraphics[width=0.75in]{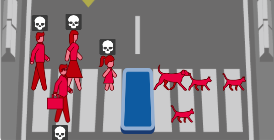}} & 649 & 0.994 & 0.115 & 0.168 \\[0.4cm]
    \parbox[c]{1em}{\includegraphics[width=0.75in]{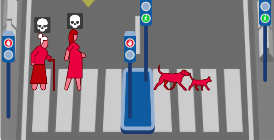}} & 924 & 0.051 & 0.591 & 0.258 \\[0.4cm]
    \parbox[c]{1em}{\includegraphics[width=0.75in]{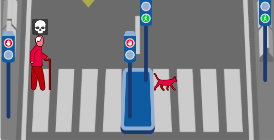}} & 2671 & 0.292 & 0.760 & 0.346 \\[0.4cm]
    \parbox[c]{1em}{\includegraphics[width=0.75in]{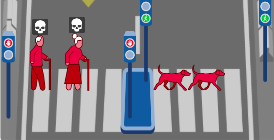}} & 146 & 0.274 & 0.736 & 0.349 \\[0.4cm]
    \parbox[c]{1em}{\includegraphics[width=0.75in]{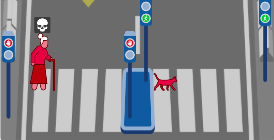}} & 2589 & 0.287 & 0.741 & 0.338 \\
    \bottomrule
    \end{tabular}
    \label{tab:datresid}
\end{table}

Table \ref{tab:nnresid} reports the largest \textit{smoothed} residuals, \textit{i.e.} the largest differences between the choice model and the neural network. We suggest that these residuals reflect the `true residuals' better than the data. In these dilemmas, participants must decide whether the car should stay and kill the illegally crossing human or swerve and hit the legally crossing animal. Most participants chose to swerve, and the neural network correctly predicted this result. However, the `Hybrid' choice model often predicted the opposite. Looking at its coefficients, we can understand why: there was a penalty for both illegally crossing and swerving, and the sum of those penalties outweighed the utility differences between the human and the animal. We clustered those dilemmas as humans-versus-animal dilemmas, and it seemed that, in these instances, humans should be saved regardless of their crossing signal status and relationship to the side of the car. This represented a deontological principle, a moral rule independent of the consequences of the action \cite{sep-ethics-deontological}. Thus, while our `Hybrid' choice model only used two deontological principles, we added a third for future iterations: if a given dilemma requires choosing between humans or animals, humans should be preferentially saved. This feature would have been difficult to justify when looking at the residuals from the data, because the largest residual there actually exhibited a strong effect in the opposite direction. Going down the list of smoothed residuals, we are able to cluster another group of dilemmas with high errors and conducted a similar analysis shown in Supplementary Table S1. Most salient to us in those dilemmas was an age gradient. Similar to above, future iterations of our model incorporated a deontological principle explicitly favoring the young in old-versus-young dilemmas.

\begin{table}[!htb]
    \centering
    \caption{Biggest Differences Between Choice Model and Neural Network (proportions show observed or predicted proportion killing left side).}
    \begin{tabular}{p{0.75in}cccc}
    \toprule
    \parbox[c]{1em}{\includegraphics[width=0.75in]{Car.png}} & \textbf{N} & \textbf{Data}    & \textbf{Choice Model}  & \textbf{Neural Network}  \\[0.2cm]
    \midrule
    \parbox[c]{1em}{\includegraphics[width=0.75in]{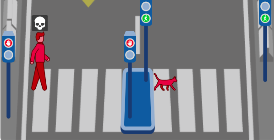}} & 2541 & 0.301 & 0.699 & 0.272 \\[0.4cm]
    \parbox[c]{1em}{\includegraphics[width=0.75in]{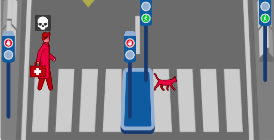}} & 2541 & 0.249 & 0.662 & 0.239 \\[0.4cm]
    \parbox[c]{1em}{\includegraphics[width=0.75in]{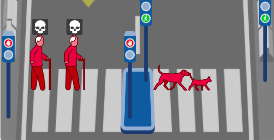}} & 153 & 0.366 & 0.746 & 0.326 \\[0.4cm]
    \parbox[c]{1em}{\includegraphics[width=0.75in]{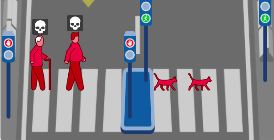}} & 146 & 0.370 & 0.715 & 0.296 \\[0.4cm]
    \parbox[c]{1em}{\includegraphics[width=0.75in]{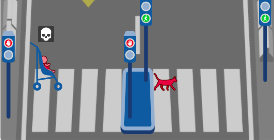}} & 2561 & 0.195 & 0.637 & 0.220 \\[0.4cm]
    \bottomrule
    \end{tabular}
    \label{tab:nnresid}
\end{table}

\subsubsection{Incorporating Axes of Differences} 

Humans versus animals and old versus young were two of six `axes of differences' the Moral Machine researchers explicitly manipulated in their experiment, the other four being fat versus fit, more versus less, male versus female, and high status versus low status. While these axes were not explicitly revealed to the participant, the residuals we identified suggested participants were sensitive to them. We incorporated these six new features as additional deontological principles into our `Hybrid' choice model and plotted the results in the bottom panel of Figure \ref{fig:initmetrics}. The new choice model, `Hybrid + Axes' had a significantly better accuracy than the `Hybrid' model, demonstrating that we were able to build a better predictive model of moral judgment while retaining interpretability and explanatory power. Furthermore, we added these axes as inputs into the neural network to create `Neural Network + Axes.' This model outperformed the original network at smaller dataset sizes but became seemingly identical at larger ones, suggesting that the original network could construct these axes once there was sufficient data. These axes were at least as complex as twenty-way interactions. 

This human part of identifying features from residuals is important in generating explanatory insights of human behavior. First, it allows the researcher to connect the new features with past research. For example, the `axes of difference' we found are reminiscent of work by Tversky regarding `features of similarity' \cite{tversky1977features}. Second, this manual step helps ensure that the researcher is incorporating psychologically meaningful features rather than spurious information. For example, Zech and colleagues \cite{zech2018confounding} found that machine learning models were overfitting to hospital-specific information in a training set of medical images, rather than validly approximating the true functional mapping between the images and diagnoses. A human-led featurization step as we propose would help ensure that the new features for the simple, interpretable model do not reflect this spurious information.

Despite the initial success in increasing accuracy after the first iteration, the model-building process still displayed a potential for improvement (as indicated by the AUC curve), and thus we conducted more iterations of our loop. Using the smoothed residuals from the second iteration, we identified axes not explicitly manipulated by the researchers, such as Pregnant Women and Doctors versus other humans, and split previous axes into sub-axes (\textit{e.g.,} young versus old was split into young versus adult, adult versus old, and young versus old). The third and fourth iterations modeled two-way and three-way conjunctive features between the axes of differences, the crossing signals, and the intervention status (\textit{e.g.} a car headed towards illegally crossing humans in a humans-versus-animals dilemma). 

\begin{table}[!htb]
    \centering
    \caption{Comparison of Model Fit under Different Metrics}
    \label{table:finmetrics}
    \begin{tabular}{lccc}
    \toprule
    \bf{Model Type} & \bf{Accuracy} & \bf{AUC} & \bf{AIC} \\ 
    \midrule
    % Equal Weight  & 0.617  & 0.628 & 1.305 \\
    % Animals vs. People & 0.681 & 0.701 & 1.239\\
    Deontological  & 0.630  & 0.631 & 1.303 \\
    Utilitarian   & 0.719  & 0.779 & 1.161 \\
    Hybrid  & 0.756  & 0.814 & 1.052 \\
    % Bayesian Brute-Force Interaction  & 0.757  & 0.816 & 1.058 \\
    Hybrid + Axes (Iteration 1)  & 0.760  & 0.823 & 1.021 \\
    Additional Axes (Iteration 2)  & 0.764  & 0.825 & 1.019  \\
    Two-Way Conjunctions (Iteration 3)  & 0.764  & 0.829 & 1.003  \\
    Three-Way Conjunctions (Iteration 4)  & 0.768  & 0.830 & 0.999 \\
    Neural Network  & 0.768 & 0.833 & 0.999 \\ 
    Empirical Upper Bound  & 0.804 & 0.890 & N/A \\ 
    \bottomrule
    \end{tabular}
\end{table}

Table \ref{table:finmetrics} displays the final results of our model-building process. It is up to the modeler to decide when to stop the process, and in this case study, we stopped when the metrics between the new choice model and the neural network were maximally close. Supplementary Tables S2 to S7 report the largest smoothed and raw residuals for the later iterations. The features we identified at these later iterations reflect more subtle and complicated principles. While there was conceptual overlap between the largest smoothed residuals and raw residuals for the first iteration, the gap seems to grow at the later iterations, in which the larger raw residuals seem to be very different than the largest smoothed residuals. Our resulting model predicted human decisions with an accuracy comparable to the neural network and was entirely interpretable (all features and their weights are outlined in Supplementary Table S8).  The table also shows the maximum possible accuracy when using the aggregate data to predict the choice for every given dilemma via a table lookup algorithm (\textit{i.e.} if $90\%$ of participants in a given dilemma chose to swerve, the empirical prediction for that dilemma will be $90\%$; as a result, it should be noted that the performance of this `model' was not calculated out-of-sample, while all the other models were).

\subsection{Empirical Results}

SRM is a form of exploratory data analysis. Such methods have the vulnerability of overfitting to data and thus need to be complemented with confirmatory data analysis techniques \cite{tukey1980we}. We identified and empirically validated interesting effects from three iterations of SRM. First, regarding a new axis of difference, we found convincing evidence that participants excluded criminals from moral protections afforded to other human agents. We previously discussed the need to incorporate a deontological principle in humans-versus-animals dilemmas that prefers saving the human side. While doing this increased the model's overall predictive power, our model started to err on a subclass of other dilemmas: criminals versus animals. In order to build a better model of human moral judgment, we had to introduce a separate criminals-versus-animals feature, thus dehumanizing criminals in the eyes of our model.

Second, we were able to identify an intuitive interaction between kids and an illegal crossing status. Consider two dilemmas (illustrated in Supplementary Figure S2) where in the first, the participant must choose between saving an old woman or a girl and in the second, the participant must choose between saving either an old woman and a woman, or a girl and a woman. Rational choice models are based on a linear utility function and would consider these dilemmas to be treated equivalently, but the Moral Machine data and the neural network revealed that participants did not always do so. Rather, participants treated the dilemmas as equivalent when the side with children was crossing legally or if there was an absence of a crossing signal, but not when the side with children was illegally crossing. In the latter cases, the side with children in the second dilemma (\textit{i.e.,} with an adult) was penalized more than the corresponding side in the first dilemma.

Lastly, there was an intriguing asymmetric interaction between car side and crossing signal status in both male-versus-female dilemmas and fat-versus-fit dilemmas. Here, when the car was headed towards the higher-valued individual (\textit{i.e.,} the female or the athlete) in the absence of a crossing signal, the probability of saving the individual was roughly halfway between the probability of saving them when they were legally crossing and the probability of saving them when they were illegally crossing. However, this relationship did not hold when the car was headed towards the lower-valued individual. Rather, in those cases, the probability of saving the individual was significantly lower than the halfway point and close to the probability of saving them when they were illegally crossing. Intuitively, lower-valued individuals aren't given the ``benefit of the doubt'' when their crossing legality is ambiguous.

We ran three preregistered experiments on Amazon's Mechanical Turk in order to replicate and confirm these effects revealed by SRM.

\subsection{Experiment 1: Criminal Dehumanization}

In this experiment, participants chose between saving a human and a dog. We varied the car side (dog, human), type of human (criminal, homeless man, old man, adult man), and crossing signal status (legally crossing, illegally crossing, N/A) for a total of twenty-four dilemmas. Each participant saw four of these twenty-four dilemmas. We calculated the percentage of participants that chose to save the human over the dog in every dilemma. For each car side and crossing signal combination, we conducted a Chi-squared test determining whether participants chose to save criminals less than each of the other three humans. This resulted in eighteen separate Chi-squared analyses, and for these eighteen analyses, criminals were saved at a rate between $11\%$ to $28\%$ less than the other human agents. All analyses were significant at the $\alpha = 0.05$ level, and seventeen of the eighteen were significant at the $\alpha = 0.001$ level. Graphical results are displayed in Figure \ref{fig:exp1} and tabular results are represented in Supplementary Table S9. The original Moral Machine results are reported in Supplementary Table S10.

\begin{figure}[!htb]
    \centering
    \includegraphics[width=\linewidth]{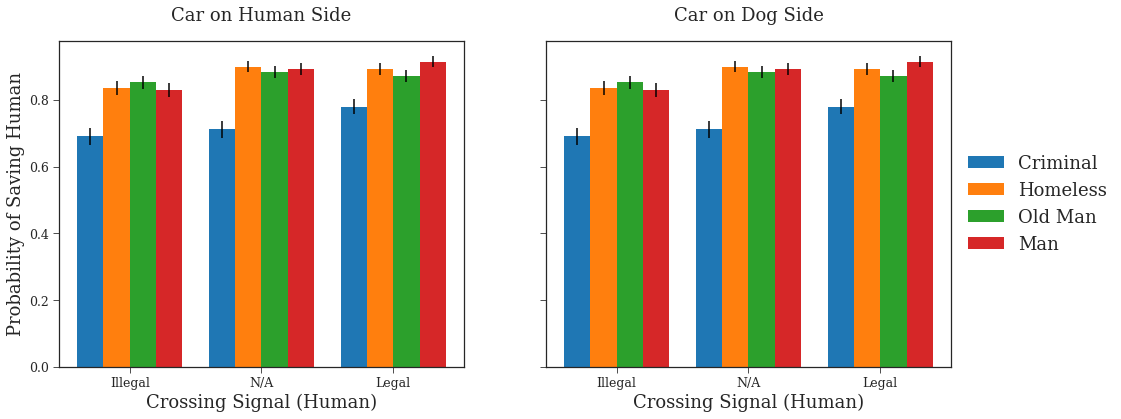}
    \caption{Dehumanization of Criminals. When pitted against dogs, participants save criminals at a significantly lower rate that other human agents.}
    \label{fig:exp1}
\end{figure}

Our results in Experiment 1 suggest that criminals are excluded from certain protections most humans are given, namely preferring to save them compared to dogs. These findings are consistent with a long line of work in sociology and psychology suggesting criminals are treated as a lower class of individuals than others in society when it comes to evaluating their status as a human being \cite{jahoda1999images,viki2012role,bastian2013roles,haslam2014dehumanization}. Opotow et. al \cite{opotow1990moral} proposed that dehumanization is a form of moral exclusion in which a victim can lose their entitlement to compassion. Besides moral exclusion, other potential frameworks to understand participants' behavior may be through retributive justice \cite{darley2003psychology,witvliet2008retributive} and standard consequentialist reasoning. We believe both of these factors were also present in this paradigm, but that they were already taken into account in our choice model as the inferred weight given to criminals. The moral exclusion argument is supported by the fact that incorporating a humans-versus-animals principle was an important predictor of Moral Machine behavior, but that we had to specifically remove this label from situations that pitted criminals versus animals. Since these axes of differences were derived from the features of the agents \cite{kim2018computational}, our modeling suggests that participants did not honor the `human' feature for criminals.

\subsection{Experiment 2: Age of Responsibility}

In this experiment, participants either chose between saving a child or an old adult or they chose between saving a child and an adult versus an old adult and an adult. We varied car side (child, old adult), crossing signal condition (legally crossing, illegally crossing, N/A), and sex (male, female) for a total of twenty-four stimuli. Each participant saw six of the twenty-four dilemmas. We aggregated responses for all dilemmas in order to calculate the percentage of participants that chose to save the young side. For each car side, sex, and crossing signal combination, we conducted a Chi-squared analysis comparing the percentage that saved the young side in a child versus old adult dilemma to the percentage that saved the young side in a child and adult versus old adult and adult dilemma. Of these twelve analyses, we hypothesized four would be significant while the other eight would not be. Specifically, we hypothesized that the analyses where the young side was crossing illegally would be significantly different but that the dilemmas in the other crossing signal conditions would not be. Three of the four hypothesized significant effects were significant at the $\alpha = 0.05$ level, while seven of the eight hypothesized null effects were not significant at the $\alpha = 0.05$ level. Results are graphically represented in Figure \ref{fig:exp2} and reported in Supplementary Table S11. The original Moral Machine results are reported in Supplementary Table S12.

\begin{figure}[!htb]
    \centering
    \subfloat{\includegraphics[width=\linewidth]{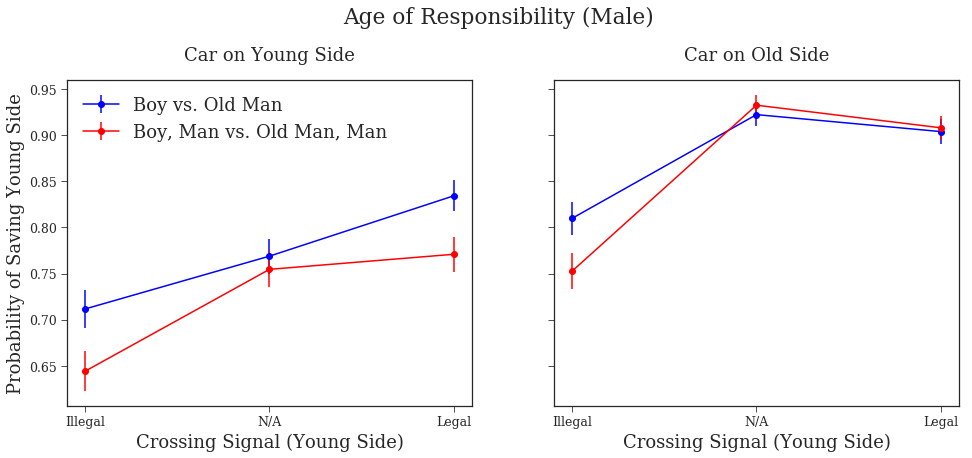}}  \hfill
    \subfloat{\includegraphics[width=\linewidth]{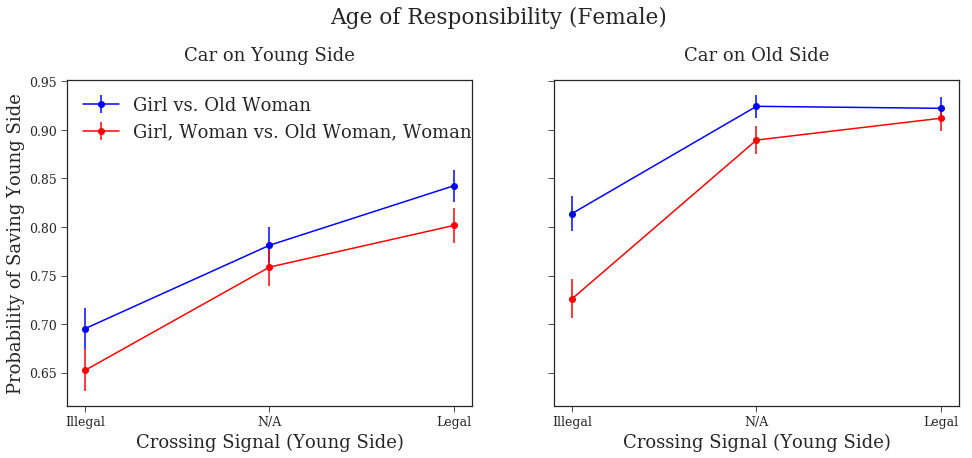}}
    \caption{Age of Responsibility. Graphs demonstrate the differences in participants' judgments when deciding between a child and an old adult versus when deciding between a child and an adult versus an adult and an old adult. The dilemmas are roughly equivalent when the side with children are either crossing legally or when there is absence of a crossing signal, but not when they are crossing illegally.}
    \label{fig:exp2}
\end{figure}

The results from Experiment 2 suggest children are given a privileged status when assigning blame. The jurisprudential logic for the privileged status of children in the law is that children often lack the \textit{mens rea}, \textit{i.e.,} the knowledge of wrongdoing and a necessary condition for criminal conviction, when partaking in illegal activity  \cite{platt1966origins,dalby1985criminal,bandalli1998abolition}. (An intuition for why \textit{mens rea} is considered important is encapsulated by Justice Oliver Wendell Holmes Jr.'s famous quip: ``Even a dog distinguishes between being stumbled over and kicked.'' \cite{holmes1881common}) Earlier, we proposed that the negative penalty associated with crossing illegally is justified by a consensual theory of punishment \cite{nino1983consensual}, in which an individual waives their rights to being protected by the law when committing an illegal action. In our experiment, when the illegally crossing pedestrians were solely comprised of children, participants did not penalize them as much as when there was one adult.  Formally, the jurisprudential logic behind participants' decisions here would be that the children did not have the necessary \textit{mens rea} when crossing illegally and thus they did not willingly waive their rights to being protected by the law. As a result, they should not be penalized as much as adults, who presumably did have the \textit{mens rea} and thus knowingly waived their rights. Furthermore, the empirical effect is stronger when the car is on the side of the old adult, which is intuitive under the consensual theory of punishment framework as it seems more reasonable to excuse a child compared to an adult for not realizing they were crossing illegally when the car was on the opposite side.

\subsection{Experiment 3: Asymmetric Notions of Responsibility}

Each dilemma in this experiment was either a male versus female or an athlete versus a large person. We varied car side and crossing signal status, as well as age (adult, old) for the male-female dilemmas and sex for the fat-fit dilemmas, for a total of twenty-four dilemmas. Each participant only saw four of the twenty-four possible dilemmas. For each axis (\textit{i.e.,} male-female or fat-fit) and car side combination, we conducted a Chi-squared analysis comparing the percentage that saved the higher-valued individual in the absence of a crossing signal to the average of the percentages that saved in the legal and illegal crossing settings. We hypothesized that when the car was headed towards the lower-valued individuals, the proportion saved in the absence of a crossing signal condition would be significantly less than the mean of the other two crossing signal settings, while we did not think there would be a significant difference when the car was headed towards the higher-valued individuals. All four of our hypothesized significant effects were significant at the $\alpha = 0.05$  level and all four of our hypothesized null effects were not significant at this level. Results are graphically represented in Figure \ref{fig:exp3} and reported in Supplementary Table S13. The original Moral Machine results are reported in Supplementary Table S14.

\begin{figure}[!htb]
    \centering
    \subfloat{\includegraphics[width=\linewidth]{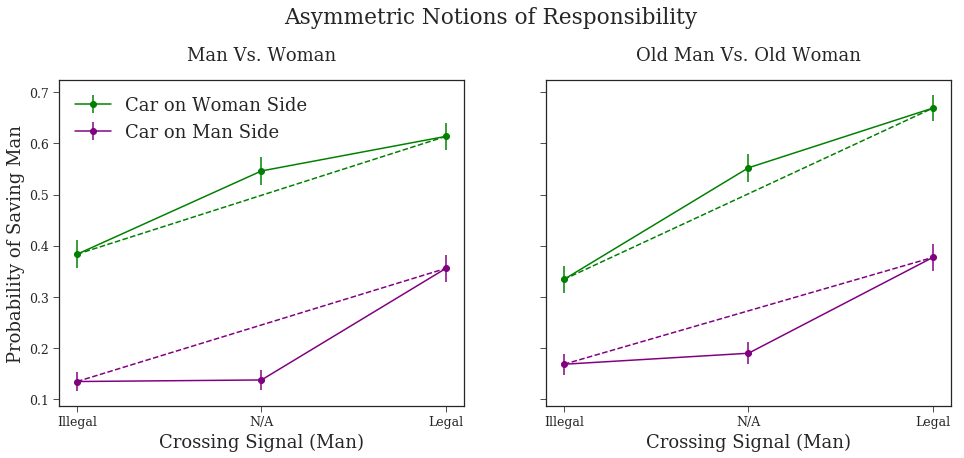}}  \hfill
    \subfloat{\includegraphics[width=\linewidth]{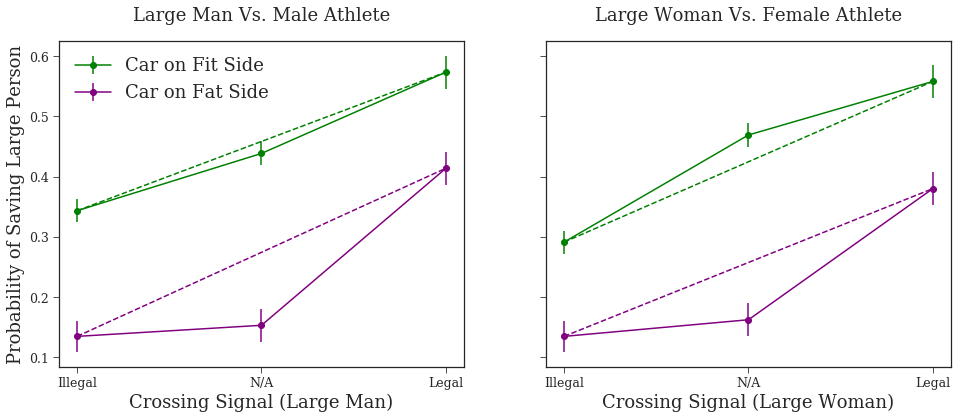}}
    \caption{Asymmetric Notions of Responsibility. The dotted line indicates the average of the legal and illegal crossing conditions. When the car is headed towards the high-valued individual, their judgments are close to that predicted by the dotted line. However, when the car is headed towards the lower-valued individual, their judgments are close to the ones in which the individual is crossing illegally.}
    \label{fig:exp3}
\end{figure}

The results in Experiment 3 demonstrated that when the car is headed towards the higher-valued individual and there is an absence of a crossing signal, the individual is treated half as if they are crossing legally and half as if they are crossing illegally. The same is not true when the car is headed towards the lower-valued individual. In those cases, the individual is treated in almost the same manner as when they are illegally crossing. One conjecture for this behavior is a form of motivated reasoning \cite{kunda1990case,alicke2000culpable,ditto2009motivated}. Participants may have started off by assuming that the pedestrian in the same lane as the car is the one at fault. However, because the participant was motivated to save the higher-valued individual, they treated the absence of a crossing signal as an ambiguity that suggested equal probability of crossing legally or illegally. Conversely, when the car is headed towards the lower-valued individual, participants may have been motivated to infer that the individual was probably crossing illegally, and thus use the fact they are in front of the car to justify this belief.

\section{Discussion}

When there is so much data in front of us, where do we even start to look? This problem is not unique to large-scale experiments. Rather, it is the problem of the scientific enterprise in general. The scientific method has offered a solution: identify the signal in the data and iteratively critique hypotheses until they are able to explain as much of the signal as possible. In this paper, we formalized this idea as an iterative loop in which we critique interpretable and theoretically-constrained psychological models with respect to a data-driven machine learning algorithm. Standard forms of exploratory data analysis critique models with respect to the data, but once the dataset is sufficiently large, a purely data-driven machine learning algorithm like a neural network can often provide a better estimate of the true underlying function than the data itself.

We illustrated this methodology in the domain of moral decision-making. Psychological models of moral reasoning are often derived from consequentialist and deontological theories in moral philosophy \cite{kant1785groundwork,bentham1789introduction}, and these theories have been extremely fruitful in motivating moral psychology research. However, it is inevitable that a highly theoretically-driven scientific program will lead to incomplete models of human behavior. By contrasting these constrained models with data-driven models, we were able to identify shortcomings and use them to build a model that is both theoretically grounded and powerfully predictive. We found that incorporating axes of differences and their interactions with other deontological principles improved the accuracy of a rational choice model of moral decision-making. We then validated three of our findings by running independent preregistered experiments. %In the remainder of the paper, we relate these findings to the moral reasoning literature and highlight the implications of our results.

Our work is conceptually similar to model compression in which a `simple' model is trained on the predictions of a complex model \cite{bucilua2006model}. However, in that line of work, simplicity is defined with respect to a runtime processing whereas in ours is defined with respect to interpretability. Both our work and theirs leverage the fact that a neural network can serve as a universal function approximator \cite{hornik1989multilayer,hartman1990layered}. They use it as their rationale to use a neural network to approximate the predictions of boosting trees, while we use it as our rationale to estimate the true underlying function. Because neural networks are the `simple' model in model compression, there is no residual analysis, and thus the majority of the work is dedicated to identifying ways to create a large dataset so that a neural network can be trained on, while the majority of our methodology is centered around residual analysis.

SRM is also similar to research by Rudin and colleagues \cite{rudin2010process,rudin2019stop}, in which the goal is to create interpretable machine learning models for high-stakes decisions. Our results in Table \ref{table:finmetrics} demonstrate that there is not necessarily a tradeoff between accuracy and interpretability, as commonly thought by many machine learning researchers. Rather, if given structured features, interpretable models can perform similarly to (and perhaps even outperform) black-box machine learning models. The methodology we propose in this paper is a systematic process for identifying and building structured features in the data.

The Moral Machine dataset proved to be a fruitful case study for Scientific Regret Minimization: rational choice models performed well, but we were still able to use a neural network to identify shortcomings once the dataset became sufficiently large. We expect that this methodology can be used in different domains, especially in mature fields (which may have unwittingly missed important systematic effects), but also in newer fields wherein the gaps between theoretically-inspired models and data-driven models remain large. Future work can extend our methodology in at least two different ways. The first is automating the identification and clustering of residuals into human-interpretable features. The second is that, while we assumed a specific functional form (\textit{i.e.,} a rational choice model) for the final model, it is plausible that this theoretical model is incorrect and thus we may need to develop a systematic way to identify the proper functional form itself.

% However, when using this methodology, it is important to use best practices in machine learning. For example, Zech and colleagues \cite{zech2018confounding} found that machine learning models were picking up on spurious information in the training set and thus were not able to perform well on different testing sets. The Moral Machine dataset was prepared in a way such that confounding variables weren't present, and this was necessary for the success of our methodology. When using this methodology in different domains, care needs to be taken to eliminate any sources of confounding, \textit{e.g.} training and testing sets should be drawn from the same distribution and evaluation on a held-out testing set should be the standard.

Lastly, on a broader note, we hope to further the development of a synergistic correspondence between psychology and data science approaches in scientific modeling \cite{rosenfeld2012combining,dwyer2018machine,peterson2018evaluating,bourgin2019cognitive}. Cognitive science famously grew out of the intersection of six different fields \cite{gardner1987mind}, but some have suggested that this revolution did not create the emergence of a new discipline \cite{lakatos1986methodology,miller2003cognitive,nunez2019}. Rather, research often proceeds independently in each contributing field. One potential reason for the lack of unification lies on a philosophical level: different scientific traditions have different epistemic values and are methodologically incommensurable \cite{kuhn1962structure}. For example, psychology prioritizes explanation while machine learning is almost exclusively focused on prediction, and their methodologies reflect these differences \cite{hofman2017prediction,yarkoni2017choosing,jolly2019flatland}. To live up to promise of the cognitive revolution, we need to truly integrate the different values and methodologies implicit in these related fields. We hope the approach in this paper offers a step in that direction.

\section{Methods}

\subsection{Mathematical Analysis and Simulations}

The proof for the result in Equation $\ref{eq:regretderivation}$ is below:

\begin{flalign*}
  & \Expect_{p(x, y)}\bigg[y - g(x)\bigg]^2&& \\
&\quad = \Expect_{p(x, y)}\bigg[\big(f(x) + \epsilon\big) - g(x)\bigg]^2&& \notag \\
&\quad = \Expect_{p(x, y)}\bigg[\big(f(x) - g(x)\big) +\epsilon\bigg]^2&& \notag \\
&\quad = \Expect_{p(x, y)}\bigg[\big(f(x) - g(x)\big)^2 +2\epsilon\big(f(x)-g(x)\big) + \epsilon^2\bigg]&& \notag \\
&\quad = \Expect_{p(x)}\bigg[\big(f(x) - g(x)\big)^2\bigg] + \Expect_{p(x)}\bigg[2\epsilon\big(f(x)-g(x)\big)\bigg] + \Expect_{p(y)}\bigg[\epsilon^2\bigg]&& \notag \\
&\quad = \Expect_{p(x)}\bigg[\big(f(x) - g(x)\big)^2\bigg] + \Expect_{p(x)}\bigg[2\epsilon\big(f(x)-g(x)\big)\bigg] + \sigma_{\epsilon}^2&& \notag \\
&\quad =\Expect_{p(x)}\bigg[\big(f(x) - g(x)\big)^2\bigg] + \sigma_{\epsilon}^2&  \notag %\text{$\Expect_{p(x)}[\epsilon] = 0$} & \notag
\end{flalign*}

The proof for the result in Equation $\ref{eq:nnregretderivation}$ is the following:
\begin{flalign*}
\begin{aligned}
& \Expect_{p(x, y)}\bigg[\hat{f}(x) - g(x)\bigg]^2 \notag \\
&\quad = \Expect_{p(x)}\bigg[\big(f(x) - g(x)\big) + \big(\hat{f}(x) - f(x)\big)\bigg]^2 \notag \\
&\quad = \Expect_{p(x)}\bigg[\big(f(x) - g(x)\big)^2 + 2\big(f(x) - g(x)\big)\big(\hat{f}(x) - f(x)\big) + \big(\hat{f}(x) - f(x)\big)^2\bigg] \notag
\end{aligned}
\end{flalign*}

For Figure \ref{fig:regretmin}, data was generated from the polynomial function $3x(x-2)^2(x+2)^2(x+1)$, and the input was uniformly sampled from the domain $[-2.5, 2.5]$ and rounded to the nearest thousandth, thus allowing for multiple samples of the same data point. Each data point had noise independently drawn from a normal distribution $\mathcal{N}(0, 10)$. The neural network used a `ReLU' activation function and had two hidden layers, the first with one hundred hidden neurons and the second with fifty hidden neurons. Ten different simulations were run for each different dataset size.

\subsection{Computational Modeling}

The neural network was trained to minimize the binary crossentropy between the model's output and human binary decisions. We conducted a grid search on the space of hyperparameters to identify the optimal settings for the network. A neural network with three 32-unit hidden layers and a `ReLU' activation function was used for all the analyses in this paper. Keras \cite{chollet2018keras} was used for training the neural networks, and the networks were optimized through Adam \cite{kingma2014adam}. Logistic regression models were trained via sci-kit learn \cite{pedregosa2011scikit}.

When calculating metrics for a given dataset size, five samples of that size were bootstrapped from the whole dataset. Each sample was split into training and testing sets. Train/test splits were based on unique dilemmas as opposed to individual judgments. There was a wide distribution of the number participant judgments per unique dilemma, and we wanted both the training and test sets to have similar distributions. Thus, in order to approximate an 80/20 split, we sorted the dilemmas by the number of judgments and binned the dilemmas into groups of five. For every bin, four were randomly assigned to the training set and the fifth was assigned to the testing set. As a result, all train/test splits were approximately, but not exactly, 80/20 splits.

\subsection{Empirical Results}

2,086 participants across twelve conditions were recruited from Amazon Mechanical Turk and paid $\$0.50$ to participate in an experiment in which they indicated their preferences in twenty-eight Moral Machine autonomous car dilemmas. The order of all twenty-eight dilemmas was randomized for each participant. Five of the twenty-eight dilemmas were attention checks. In the attention checks, participants had the option of either saving or killing everyone in the dilemma. If they chose to kill everyone more than once, they were excluded from further analysis. The experiment's preregistration called for $163$ participants per condition (twelve conditions for a total $N = 1,956$) after the exclusion criteria were applied.

Nine of the remaining twenty-three dilemmas were passengers versus pedestrian dilemmas while fourteen were the stimuli for the hypotheses. The nine passengers versus pedestrian dilemmas were included to add variation because the fourteen stimuli used for the hypotheses were all pedestrian versus pedestrian dilemmas. Answers for these dilemmas were not analyzed. Furthermore, both the nine passengers versus pedestrian dilemmas and five attention checks were kept constant across all twelve conditions. 

Because there were a total twenty-four possible stimuli for each hypothesis, Hypothesis 1 and Hypothesis 3 stimuli were split into six groups of four and allocated throughout the twelve conditions such that each group was assigned to two conditions. Hypothesis 2 stimuli were split into four groups of six and allocated such that each group was assigned to three conditions. Thus, of the fourteen dilemmas participants saw for the hypotheses, four were for Hypothesis 1, six were for Hypothesis 2, and four were for Hypothesis 3. The end result was that all Hypothesis 1 and Hypothesis 3 stimuli received $326$ judgments while all Hypothesis 2 stimuli received $489$ judgments. These sample sizes were chosen in order to achieve $95\%$ power at detecting a true effect using the Chi-squared proportion test at $\alpha = 0.05$. Effect sizes were estimated using results from the Moral Machine dataset. It should be noted that our procedure was different than the original Moral Machine paradigm, which asked participants thirteen dilemmas and operated over a wider range of experimental manipulations.

Experiments were coded using the jsPsych software package \cite{de2015jspsych} and the interface with Amazon Mechanical Turk was provided with psiTurk \cite{gureckis2016psiturk}. The dilemmas were created using the `Design' feature on the Moral Machine website.

Data from the experiments and the analysis script for the figures in this paper are uploaded at https://osf.io/25w3v/?view\_only=b02f56f76f7648768ce3addd82f16abd. The preregistration can also be accessed from there.

\section{Acknowledgements}
We thank Edmond Awad for providing guidance on navigating the Moral Machine dataset. M.A. is supported by the National Defense Science and Engineering Graduate Fellowship Program.

\bibliography{refs}
\bibliographystyle{apacite}

\clearpage
\section{Supplementary Material}
\beginsupplement

\subsection{Bayesian Feature Selection}

To simulate an alternative approach towards exploratory data analysis, we conducted a form of Bayesian feature selection \cite{mitchell1988bayesian}. We trained a Bayesian logistic regression model with all `Hybrid' model features and their two- and three-way interactions. Each weight was given a prior of a Gaussian distribution with mean $0$ and standard deviation $0.1$. Once this model was trained, all features in which a weight of $0$ was located in its 95$\%$ credible interval were removed. We then trained this new model, and repeated this procedure until all features that were fit were significant. (More computationally intensive variable selection procedures, such as marginal likelihood, were infeasible given the size of the dataset).  Table S15 outlines the iterations' metrics and Table S16 reports the final features and their weights.

The resulting model from this approach performed a little better than the original `Hybrid' model and far worse than our final choice model (and in fact worse than the model after our first iteration). It seems that for such an approach to rival ours, we would have needed to start off with a model that encapsulated at least all twenty-way interactions. Such a model would be even more intractable to conduct for Bayesian feature selection.

\subsection{Methods}

Due to the size of the dataset and the feature set, this model was trained through variational inference \cite{blei2017variational} rather than traditional MCMC sampling. The model was trained using a Flipout gradient estimator \cite{wen2018flipout} and optimized via Adam \cite{kingma2014adam}. Metrics were computed by taking the MAP estimate of each weight. The model was trained using the Tensorflow Probability package \cite{dillon2017tensorflow}  .

\clearpage

\begin{figure}[!htb]
    \centering
    \includegraphics[clip,width=0.5\textwidth]{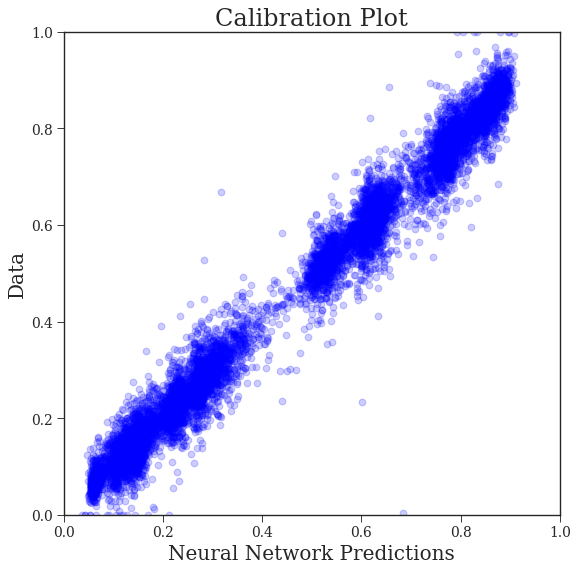}
    \caption{A calibration plot between the neural network's predictions and the aggregate dilemmas for all dilemmas over one hundred responses. We calculated a line of best fit for all dilemmas (\textit{i.e.} not just those with over hundred responses), weighting each dilemma by the number of participants that answered it. The line of best fit had a slope of 1.003 and an intercept of 0.001.}
    \label{fig:ageexample}
\end{figure}

\clearpage

\begin{table}[!htb]
    \centering
    \caption{Old vs. Young Dilemmas (proportions show observed or predicted proportion killing left side).}
    \begin{tabular}{p{1.5in}cccc}
    \toprule
    \parbox[c]{1em}{\includegraphics[width=1.5in]{Car.png}} & \textbf{N} & \textbf{Data}    & \textbf{Choice Model}  & \textbf{Neural Network}  \\[0.5cm]
    \midrule
    \parbox[c]{1em}{\includegraphics[width=1.5in]{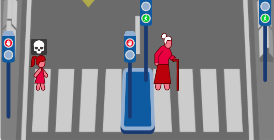}} & 11554 & 0.343 & 0.636 & 0.333 \\[0.9cm]
    \parbox[c]{1em}{\includegraphics[width=1.5in]{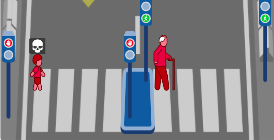}} & 11578 & 0.362 & 0.637 & 0.344 \\[0.9cm]
    \parbox[c]{1em}{\includegraphics[width=1.5in]{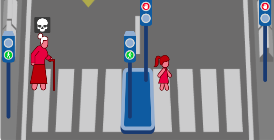}} & 7166 & 0.721 & 0.511 & 0.747 \\[0.9cm]
    \parbox[c]{1em}{\includegraphics[width=1.5in]{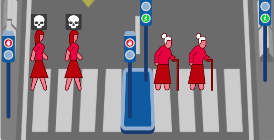}} & 5758 & 0.523 & 0.635 & 0.403 \\[0.9cm]
    \parbox[c]{1em}{\includegraphics[width=1.5in]{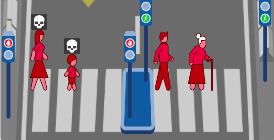}} & 5691 & 0.476 & 0.625 & 0.395 \\[0.9cm]
    \bottomrule
    \label{tab:aggyoungold}
    \end{tabular}
\end{table}

\clearpage

\begin{table}[ht]
    \centering
    \caption{Biggest Differences Between Choice Model and Data for Second Iteration}
    \begin{tabular}{p{1.5in}cccc}
    \toprule
    \parbox[c]{1em}{\includegraphics[width=1.5in]{Car.png}} & \textbf{N} & \textbf{Data}    & \textbf{Choice Model}  & \textbf{Neural Network}  \\[0.5cm]
    \midrule
    \parbox[c]{1em}{\includegraphics[width=1.5in]{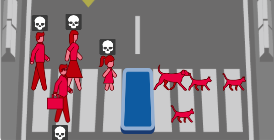}} & 649 & 0.994 & 0.164 & 0.168 \\[.9cm] 
    \parbox[c]{1em}{\includegraphics[width=1.5in]{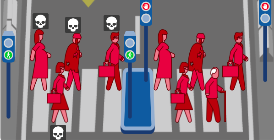}} & 1124 & 0.000 & 0.600 & 0.442\\[0.9cm]
    \parbox[c]{1em}{\includegraphics[width=1.5in]{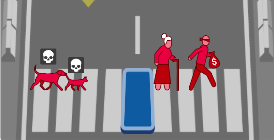}} & 1113 & 0.288 & 0.791 & 0.680 \\[0.9cm]
    \parbox[c]{1em}{\includegraphics[width=1.5in]{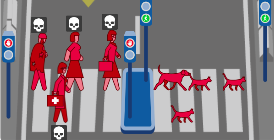}} & 890 & 0.001 & 0.396 & 0.272 \\[0.9cm]
    \parbox[c]{1em}{\includegraphics[width=1.5in]{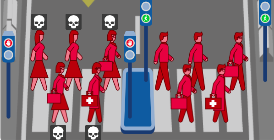}} & 365 & 0.326 & 0.709 & 0.719 \\
    \bottomrule
    \end{tabular}
    \label{tab:datresiditer2}
\end{table}

\begin{table}[ht]
    \centering
    \caption{Biggest Differences Between Choice Model and Neural Network for Second Iteration}
    \begin{tabular}{p{1.5in}cccc}
    \toprule
    \parbox[c]{1em}{\includegraphics[width=1.5in]{Car.png}} & \textbf{N} & \textbf{Data}    & \textbf{Choice Model}  & \textbf{Neural Network}  \\[0.5cm]
    \midrule
    \parbox[c]{1em}{\includegraphics[width=1.5in]{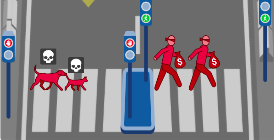}} & 130 & 0.600 & 0.869 & 0.537\\[0.9cm]
    \parbox[c]{1em}{\includegraphics[width=1.5in]{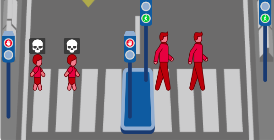}} & 1471 & 0.434 & 0.712 & 0.394 \\[0.9cm]
    \parbox[c]{1em}{\includegraphics[width=1.5in]{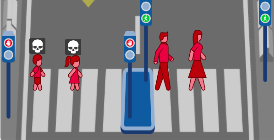}} & 2898 & 0.436 & 0.714 & 0.420 \\[0.9cm]
    \parbox[c]{1em}{\includegraphics[width=1.5in]{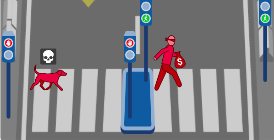}} & 3879 & 0.520 & 0.786 & 0.509 \\
    \parbox[c]{1em}{\includegraphics[width=1.5in]{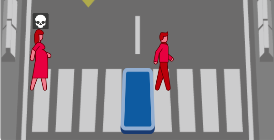}} & 3377 & 0.224 & 0.508 & 0.231 \\[0.9cm]
    \bottomrule
    \end{tabular}
    \label{tab:nnresiditer2}
\end{table}

\begin{table}[ht]
    \centering
    \caption{Biggest Differences Between Choice Model and Data for Third Iteration}
    \begin{tabular}{p{1.5in}cccc}
    \toprule
    \parbox[c]{1em}{\includegraphics[width=1.5in]{Car.png}} & \textbf{N} & \textbf{Data}    & \textbf{Choice Model}  & \textbf{Neural Network}  \\[0.5cm]
    \midrule
    \parbox[c]{1em}{\includegraphics[width=1.5in]{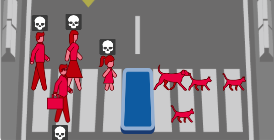}} & 649 & 0.994 & 0.155 & 0.168 \\[0.9cm] 
    \parbox[c]{1em}{\includegraphics[width=1.5in]{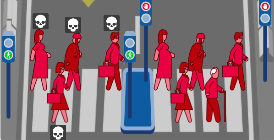}} & 1124 & 0.000 & 0.605 & 0.442\\[0.9cm]
    \parbox[c]{1em}{\includegraphics[width=1.5in]{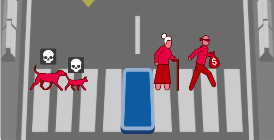}} & 1113 & 0.288 & 0.806 & 0.680 \\[0.9cm]
    \parbox[c]{1em}{\includegraphics[width=1.5in]{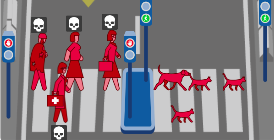}} & 890 & 0.001 & 0.693 & 0.272 \\[0.9cm]
    \parbox[c]{1em}{\includegraphics[width=1.5in]{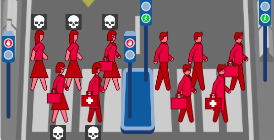}} & 365 & 0.326 & 0.709 & 0.719 \\
    \bottomrule
    \end{tabular}
    \label{tab:datresiditer3}
\end{table}

\begin{table}[ht]
    \centering
    \caption{Biggest Differences Between Choice Model and Neural Network for Third Iteration}
    \begin{tabular}{p{1.5in}cccc}
    \toprule
    \parbox[c]{1em}{\includegraphics[width=1.5in]{Car.png}} & \textbf{N} & \textbf{Data}    & \textbf{Choice Model}  & \textbf{Neural Network}  \\[0.5cm]
    \midrule
    \parbox[c]{1em}{\includegraphics[width=1.5in]{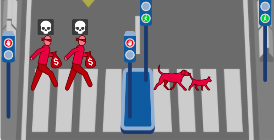}} & 162 & 0.599 & 0.835 & 0.567 \\[0.9cm] 
    \parbox[c]{1em}{\includegraphics[width=1.5in]{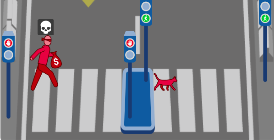}} & 2606 & 0.558 & 0.765 & 0.499\\[0.9cm]
    \parbox[c]{1em}{\includegraphics[width=1.5in]{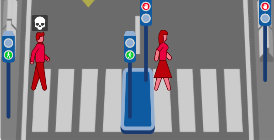}} & 8235 & 0.340 & 0.637 & 0.373 \\[0.9cm]
    \parbox[c]{1em}{\includegraphics[width=1.5in]{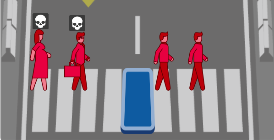}} & 175 & 0.269 & 0.541 & 0.283 \\[0.9cm]
    \parbox[c]{1em}{\includegraphics[width=1.5in]{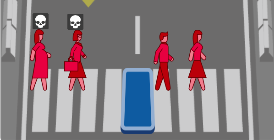}} & 359 & 0.315 & 0.539 & 0.290 \\
    \bottomrule
    \end{tabular}
    \label{tab:nnresiditer3}
\end{table}

\begin{table}[ht]
    \centering
    \caption{Biggest Differences Between Choice Model and Data for Fourth Iteration}
    \begin{tabular}{p{1.5in}cccc}
    \toprule
    \parbox[c]{1em}{\includegraphics[width=1.5in]{Car.png}} & \textbf{N} & \textbf{Data}    & \textbf{Choice Model}  & \textbf{Neural Network}  \\[0.5cm]
    \midrule
    \parbox[c]{1em}{\includegraphics[width=1.5in]{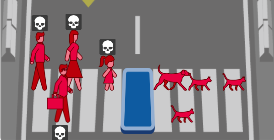}} & 649 & 0.994 & 0.147 & 0.168 \\[0.9cm] 
    \parbox[c]{1em}{\includegraphics[width=1.5in]{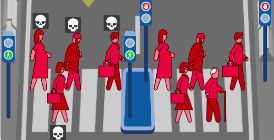}} & 1124 & 0.000 & 0.517 & 0.442\\[0.9cm]
    \parbox[c]{1em}{\includegraphics[width=1.5in]{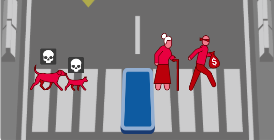}} & 1113 & 0.288 & 0.763 & 0.680 \\[0.9cm]
    \parbox[c]{1em}{\includegraphics[width=1.5in]{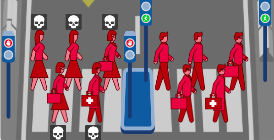}} & 365 & 0.326 & 0.766 & 0.719 \\
    \parbox[c]{1em}{\includegraphics[width=1.5in]{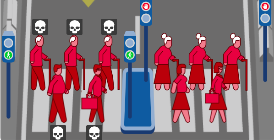}} & 187 & 0.001 & 0.427 & 0.393 \\[0.9cm]
    \bottomrule
    \end{tabular}
    \label{tab:datresiditer4}
\end{table}

\begin{table}[ht]
    \centering
    \caption{Biggest Differences Between Choice Model and Neural Network for Fourth Iteration}
    \begin{tabular}{p{1.5in}cccc}
    \toprule
    \parbox[c]{1em}{\includegraphics[width=1.5in]{Car.png}} & \textbf{N} & \textbf{Data}    & \textbf{Choice Model}  & \textbf{Neural Network}  \\[0.5cm]
    \midrule
    \parbox[c]{1em}{\includegraphics[width=1.5in]{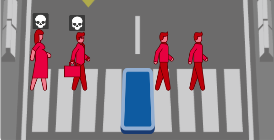}} & 175 & 0.269 & 0.560 & 0.283 \\[0.9cm] 
    \parbox[c]{1em}{\includegraphics[width=1.5in]{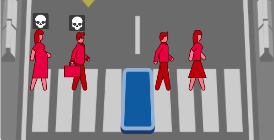}} & 326 & 0.301 & 0.564 & 0.301\\[0.9cm]
    \parbox[c]{1em}{\includegraphics[width=1.5in]{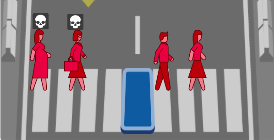}} & 359 & 0.315 & 0.552 & 0.290 \\[0.9cm]
    \parbox[c]{1em}{\includegraphics[width=1.5in]{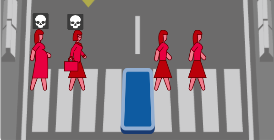}} & 172 & 0.273 & 0.556 & 0.304 \\[0.9cm]
    \parbox[c]{1em}{\includegraphics[width=1.5in]{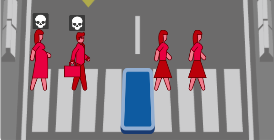}} & 159 & 0.308 & 0.568 & 0.323 \\
    \bottomrule
    \end{tabular}
    \label{tab:nnresiditer4}
\end{table}

\clearpage

\begin{longtable}{lc}
\toprule
\textbf{Feature}  & \textbf{Weight}  \\
\midrule
Man & 0.339 \\
Woman & 0.360 \\
Pregnant & 0.502 \\
Stroller & 0.537 \\
Old Man & 0.271 \\
Old Woman & 0.264 \\
Boy & 0.452 \\
Girl & 0.466 \\
Homeless & 0.208 \\
Large Woman & 0.234 \\
Large Man & 0.165 \\
Criminal & -0.093 \\
Male Executive & 0.351 \\
Female Executive & 0.371 \\
Female Athlete & 0.448 \\
Male Athlete & 0.407 \\
Female Doctor & 0.413 \\
Male Doctor & 0.427 \\
Dog & 0.333 \\
Cat & 0.285 \\
\\
Crossing Signal & 1.115 \\
\\
Car Side & -0.427 \\
\\
Humans vs. Animals & 1.034 \\
\hskip10pt Car Side on Humans & 0.828 \\
\hskip10pt Car Side on Animals & -0.207 \\
\hskip10pt Legally Crossing Humans & -0.653 \\
\hskip10pt Illegally Crossing Humans & 0.313 \\
\hskip10pt Car Side on Legally Crossing Humans & -0.303 \\
\hskip10pt Car Side on Illegally Crossing Humans & 0.189 \\
\hskip10pt Car Side on Legally Crossing Animals & -0.124 \\
\hskip10pt Car Side on Illegally Crossing Animals & 0.350 \\
Criminals vs. Animals & -0.727 \\
\hskip10pt Car Side on Criminals & -0.427 \\
\hskip10pt Car Side on Animals & 0.300 \\
\hskip10pt Legally Crossing Criminals & 0.045 \\
\hskip10pt Illegally Crossing Criminals & 0.226 \\
\hskip10pt Car Side on Legally Crossing Criminals & 0.056 \\
\hskip10pt Car Side on Illegally Crossing Criminals & 0.232 \\
\hskip10pt Car Side on Legally Crossing Animals & 0.006 \\
\hskip10pt Car Side on Illegally Crossing Animals & 0.011 \\
Pregnant vs. Other Humans & 0.338 \\
\hskip10pt Car Side on Pregnant & 0.347 \\
\hskip10pt Car Side on Other Humans & 0.009 \\
\hskip10pt Legally Crossing Pregnant & -0.109 \\
\hskip10pt Illegally Crossing Pregnant & -0.121 \\
\hskip10pt Car Side on Legally Crossing Pregnant & -0.162 \\
\hskip10pt Car Side on Illegally Crossing Pregnant & -0.068 \\
\hskip10pt Car Side on Legally Crossing Other Humans & 0.053 \\
\hskip10pt Car Side on Illegally Crossing Other Humans & -0.053 \\
Pregnant \& Doctor vs. Other Humans & 0.490 \\
\hskip10pt Car Side on Pregnant \& Doctor & 0.699 \\
\hskip10pt Car Side on Other Humans & 0.208 \\
\hskip10pt Legally Crossing Pregnant \& Doctor & -0.330 \\
\hskip10pt Illegally Crossing Pregnant \& Doctor & -0.275 \\
\hskip10pt Car Side on Legally Crossing Pregnant \& Doctor & -0.358 \\
\hskip10pt Car Side on Illegally Crossing Pregnant \& Doctor & -0.363 \\
\hskip10pt Car Side on Legally Crossing Other Humans & -0.087 \\
\hskip10pt Car Side on Illegally Crossing Other Humans & -0.028 \\
Executive \& Doctor vs. Other Humans & 0.150 \\
\hskip10pt Car Side on Executive \& Doctor & 0.056 \\
\hskip10pt Car Side on Other Humans & -0.094 \\
\hskip10pt Legally Crossing Executive \& Doctor & -0.059 \\
\hskip10pt Illegally Crossing Executive \& Doctor & -0.070 \\
\hskip10pt Car Side on Legally Crossing Executive \& Doctor & 0.002 \\
\hskip10pt Car Side on Illegally Crossing Executive \& Doctor & -0.012 \\
\hskip10pt Car Side on Legally Crossing Other Humans & 0.059 \\
\hskip10pt Car Side on Illegally Crossing Other Humans & 0.061 \\
Doctor vs. Other Humans & 0.324 \\
\hskip10pt Car Side on Doctor & 0.311 \\
\hskip10pt Car Side on Other Humans & -0.013 \\
\hskip10pt Legally Crossing Doctor & -0.146 \\
\hskip10pt Illegally Crossing Doctor & -0.195 \\
\hskip10pt Car Side on Legally Crossing Doctor & -0.177 \\
\hskip10pt Car Side on Illegally Crossing Doctor & -0.208 \\
\hskip10pt Car Side on Legally Crossing Other Humans & -0.013 \\
\hskip10pt Car Side on Illegally Crossing Other Humans & -0.032 \\
Old vs. Young & -0.402 \\
\hskip10pt Car Side on Old & -0.680 \\
\hskip10pt Car Side on Young & -0.277 \\
\hskip10pt Legally Crossing Old & 0.221 \\
\hskip10pt Illegally Crossing Old & 0.252 \\
\hskip10pt Car Side on Legally Crossing Old & 0.408 \\
\hskip10pt Car Side on Illegally Crossing Old & 0.446 \\
\hskip10pt Car Side on Legally Crossing Young & 0.194 \\
\hskip10pt Car Side on Illegally Crossing Young & 0.187 \\
Adult vs. Young & -0.150 \\
\hskip10pt Car Side on Adult & -0.457 \\
\hskip10pt Car Side on Young & -0.307 \\
\hskip10pt Legally Crossing Adult & -0.007 \\
\hskip10pt Illegally Crossing Adult & 0.053 \\
\hskip10pt Car Side on Legally Crossing Adult & 0.248 \\
\hskip10pt Car Side on Illegally Crossing Adult & 0.336 \\
\hskip10pt Car Side on Legally Crossing Young & 0.283 \\
\hskip10pt Car Side on Illegally Crossing Young & 0.255 \\
Old vs. Adult \& Young & -0.390 \\
\hskip10pt Car Side on Old & -0.722 \\
\hskip10pt Car Side on Adult \& Young & -0.332 \\
\hskip10pt Legally Crossing Old & 0.223 \\
\hskip10pt Illegally Crossing Old & 0.233 \\
\hskip10pt Car Side on Legally Crossing Old & 0.441 \\
\hskip10pt Car Side on Illegally Crossing Old & 0.500 \\
\hskip10pt Car Side on Legally Crossing Adult \& Young & 0.267 \\
\hskip10pt Car Side on Illegally Crossing Adult \& Young & 0.218 \\
Old \& Adult vs. Young & -0.207 \\
\hskip10pt Car Side on Old \& Adult & -0.471 \\
\hskip10pt Car Side on Young & -0.264 \\
\hskip10pt Legally Crossing Old \& Adult & 0.099 \\
\hskip10pt Illegally Crossing Old \& Adult & 0.223 \\
\hskip10pt Car Side on Legally Crossing Old \& Adult & 0.299 \\
\hskip10pt Car Side on Illegally Crossing Old \& Adult & 0.421 \\
\hskip10pt Car Side on Legally Crossing Young & 0.198 \\
\hskip10pt Car Side on Illegally Crossing Young & 0.200 \\
Old vs. Adult & -0.309 \\
\hskip10pt Car Side on Old & -0.714 \\
\hskip10pt Car Side on Adult & -0.405 \\
\hskip10pt Legally Crossing Old & 0.295 \\
\hskip10pt Illegally Crossing Old & 0.199 \\
\hskip10pt Car Side on Legally Crossing Old & 0.464 \\
\hskip10pt Car Side on Illegally Crossing Old & 0.465 \\
\hskip10pt Car Side on Legally Crossing Adult & 0.266 \\
\hskip10pt Car Side on Illegally Crossing Adult & 0.169 \\
Old \& Adult vs. Adult \& Young & -0.663 \\
\hskip10pt Car Side on Old \& Adult & 0.145 \\
\hskip10pt Car Side on Adult \& Young & 0.808 \\
\hskip10pt Legally Crossing Old \& Adult & 0.056 \\
\hskip10pt Illegally Crossing Old \& Adult & 0.314 \\
\hskip10pt Car Side on Legally Crossing Old \& Adult & 0.026 \\
\hskip10pt Car Side on Illegally Crossing Old \& Adult & -0.044 \\
\hskip10pt Car Side on Legally Crossing Adult \& Young & -0.358 \\
\hskip10pt Car Side on Illegally Crossing Adult \& Young & -0.031 \\
All Young vs. Other Humans & 0.104 \\
\hskip10pt Car Side on All Young & -0.055 \\
\hskip10pt Car Side on Other Humans & -0.159 \\
\hskip10pt Legally Crossing All Young & 0.022 \\
\hskip10pt Illegally Crossing All Young & 0.015 \\
\hskip10pt Car Side on Legally Crossing All Young & 0.062 \\
\hskip10pt Car Side on Illegally Crossing All Young & 0.041 \\
\hskip10pt Car Side on Legally Crossing Other Humans & 0.026 \\
\hskip10pt Car Side on Illegally Crossing Other Humans & 0.040 \\
Male vs. Female & -0.373 \\
\hskip10pt Car Side on Male & -0.353 \\
\hskip10pt Car Side on Female & 0.020 \\
\hskip10pt Legally Crossing Male & 0.204 \\
\hskip10pt Illegally Crossing Male & 0.159 \\
\hskip10pt Car Side on Legally Crossing Male & 0.440 \\
\hskip10pt Car Side on Illegally Crossing Male & 0.317 \\
\hskip10pt Car Side on Legally Crossing Female & 0.158 \\
\hskip10pt Car Side on Illegally Crossing Female & 0.236 \\
Homeless vs. Other Humans & -0.320 \\
\hskip10pt Car Side on Homeless & -0.146 \\
\hskip10pt Car Side on Other Humans & 0.174 \\
\hskip10pt Legally Crossing Homeless & 0.139 \\
\hskip10pt Illegally Crossing Homeless & 0.134 \\
\hskip10pt Car Side on Legally Crossing Homeless & 0.196 \\
\hskip10pt Car Side on Illegally Crossing Homeless & 0.081 \\
\hskip10pt Car Side on Legally Crossing Other Humans & -0.053 \\
\hskip10pt Car Side on Illegally Crossing Other Humans & 0.057 \\
Executives vs. Homeless & 0.001 \\
\hskip10pt Car Side on Executives & -0.149 \\
\hskip10pt Car Side on Homeless & -0.150 \\
\hskip10pt Legally Crossing Executives & -0.029 \\
\hskip10pt Illegally Crossing Executives & 0.024 \\
\hskip10pt Car Side on Legally Crossing Executives & 0.214 \\
\hskip10pt Car Side on Illegally Crossing Executives & 0.187 \\
\hskip10pt Car Side on Legally Crossing Homeless & 0.163 \\
\hskip10pt Car Side on Illegally Crossing Homeless & 0.243 \\
Executives vs. Adult & 0.092 \\
\hskip10pt Car Side on Executives & -0.125 \\
\hskip10pt Car Side on Adult & -0.217 \\
\hskip10pt Legally Crossing Executives & 0.042 \\
\hskip10pt Illegally Crossing Executives & -0.214 \\
\hskip10pt Car Side on Legally Crossing Executives & 0.192 \\
\hskip10pt Car Side on Illegally Crossing Executives & 0.044 \\
\hskip10pt Car Side on Legally Crossing Adult & 0.258 \\
\hskip10pt Car Side on Illegally Crossing Adult & 0.150 \\
More vs. Less & 0.800 \\
\hskip10pt Car Side on More & 0.498 \\
\hskip10pt Car Side on Less & -0.302 \\
\hskip10pt Legally Crossing More & -0.617 \\
\hskip10pt Illegally Crossing More & -0.375 \\
\hskip10pt Car Side on Legally Crossing More & -0.200 \\
\hskip10pt Car Side on Illegally Crossing More & 0.007 \\
\hskip10pt Car Side on Legally Crossing Less & 0.383 \\
\hskip10pt Car Side on Illegally Crossing Less & 0.417 \\
Fat vs. Fit & -0.293 \\
\hskip10pt Car Side on Fat & -0.313 \\
\hskip10pt Car Side on Fit & -0.021 \\
\hskip10pt Legally Crossing Fat & 0.305 \\
\hskip10pt Illegally Crossing Fat & 0.194 \\
\hskip10pt Car Side on Legally Crossing Fat & 0.539 \\
\hskip10pt Car Side on Illegally Crossing Fat & 0.360 \\
\hskip10pt Car Side on Legally Crossing Fit & 0.166 \\
\hskip10pt Car Side on Illegally Crossing Fit & 0.233 \\
Fat vs. Adult & -0.392 \\
\hskip10pt Car Side on Fat & -0.479 \\
\hskip10pt Car Side on Adult & -0.086 \\
\hskip10pt Legally Crossing Fat & 0.377 \\
\hskip10pt Illegally Crossing Fat & 0.131 \\
\hskip10pt Car Side on Legally Crossing Fat & 0.619 \\
\hskip10pt Car Side on Illegally Crossing Fat & 0.366 \\
\hskip10pt Car Side on Legally Crossing Adult & 0.235 \\
\hskip10pt Car Side on Illegally Crossing Adult & 0.241 \\
Adult vs. Fit & -0.071 \\
\hskip10pt Car Side on Adult & -0.271 \\
\hskip10pt Car Side on Fit & -0.200 \\
\hskip10pt Legally Crossing Adult & 0.204 \\
\hskip10pt Illegally Crossing Adult & -0.063 \\
\hskip10pt Car Side on Legally Crossing Adult & 0.462 \\
\hskip10pt Car Side on Illegally Crossing Adult & 0.306 \\
\hskip10pt Car Side on Legally Crossing Fit & 0.369 \\
\hskip10pt Car Side on Illegally Crossing Fit & 0.258 \\ \\ \\
\caption{Features and Weights for Final Choice Model} % needs to go inside longtable 
\end{longtable}

\clearpage

\begin{figure}[!htb]
    \centering
    \subfloat[]{\includegraphics[trim={70 90 70 50},clip,width=0.5\textwidth]{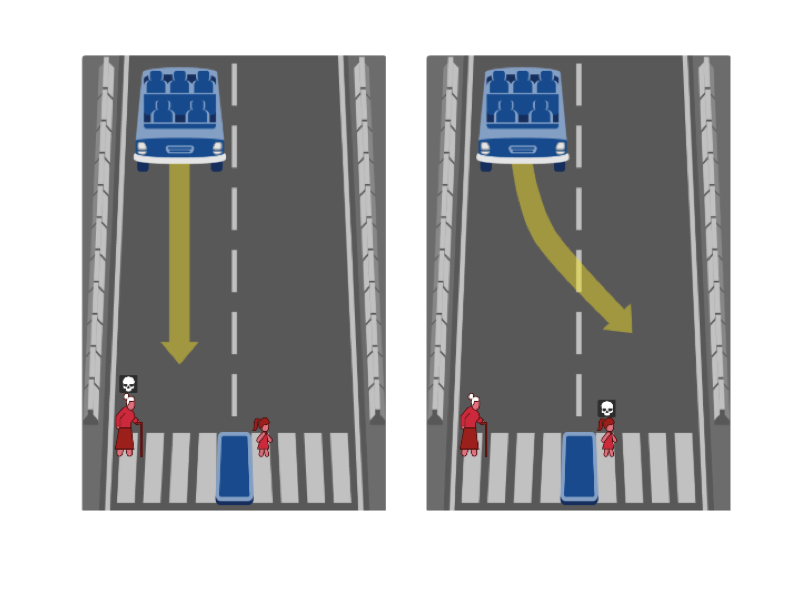}} \\
    \subfloat[]{\includegraphics[trim={70 90 70 50},clip,width=0.5\textwidth]{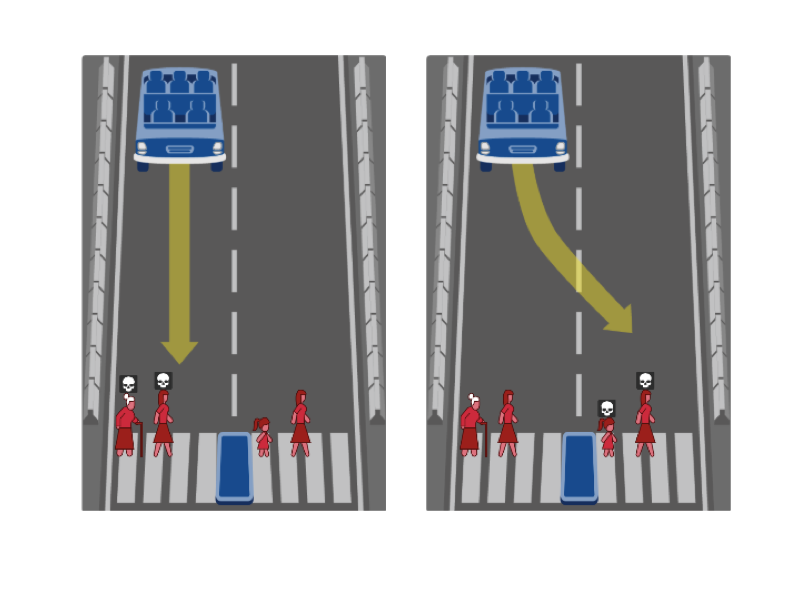}} \hfill
    \caption{Two Moral Machine dilemmas that demonstrate an age gradient. Rational choice models treat these dilemmas equivalently, but the data indicated that participants do not do so when the side with children is illegally crossing.}
    \label{fig:ageexample}
\end{figure}

\begin{table}[ht]
    \centering
    \subfloat{
    \begin{tabular}{cccccc}
    \toprule
    \textbf{Car Side} & \textbf{Signal (Human)} & \textbf{$\%$ Save Criminal} & \textbf{$\%$ Save Homeless}    & \textbf{$p$-value} & \textbf{Prediction}         \\ 
    \midrule 
    Human & Legal & 0.65 & 0.88 & $p < .001$ & Significant\\
    Human & N/A & 0.68 & 0.84 & $p < .001$ & Significant\\
    Human & Illegal & 0.63 & 0.79 & $p < .001$ & Significant\\
    Dog & Legal & 0.78 & 0.89 & $p < .001$ & Significant\\
    Dog & N/A & 0.71 & 0.90 & $p < .001$ & Significant\\
    Dog & Illegal & 0.69 & 0.83 & $p < .001$ & Significant\\
    \bottomrule
    \end{tabular}}
    \qquad
    \subfloat{
    \begin{tabular}{cccccc}
    \toprule
    \textbf{Car Side} & \textbf{Signal (Human)} & \textbf{$\%$ Save Criminal} & \textbf{$\%$ Save Old Man}    & \textbf{$p$-value} & \textbf{Prediction}         \\ 
    \midrule 
    Human & Legal & 0.65 & 0.87 & $p < .001$ & Significant\\
    Human & N/A & 0.68 & 0.82 & $p < .001$ & Significant\\
    Human & Illegal & 0.63 & 0.81 & $p < .001$ & Significant\\
    Dog & Legal & 0.78 & 0.87 & $p = .002$ & Significant\\
    Dog & N/A & 0.71 & 0.88 & $p < .001$ & Significant\\
    Dog & Illegal & 0.69 & 0.85 & $p < .001$ & Significant\\
    \bottomrule
    \end{tabular}}
    \qquad
    \subfloat{
    \begin{tabular}{cccccc}
    \toprule
    \textbf{Car Side} & \textbf{Signal (Human)} & \textbf{$\%$ Save Criminal} & \textbf{$\%$ Save Man}    & \textbf{$p$-value} & \textbf{Prediction}       \\ 
    \midrule 
    Human & Legal & 0.65 & 0.89 & $p < .001$ & Significant\\
    Human & N/A & 0.68 & 0.85 & $p < .001$ & Significant\\
    Human & Illegal & 0.63 & 0.81 & $p < .001$ & Significant\\
    Dog & Legal & 0.78 & 0.91 & $p < .001$ & Significant\\
    Dog & N/A & 0.71 & 0.89 & $p < .001$ & Significant \\
    Dog & Illegal & 0.69 & 0.83 & $p < .001$ & Significant\\
    \bottomrule
    \end{tabular}}
    \caption{Results from Experiment 1 comparing the percentage of participants that save criminals versus dogs and the percentage of participants that save other humans versus dogs. We used a $\chi^2$ analysis between the proportions, where $N = 326$ and $df = 1$.}
    \label{tab:exp1}
\end{table}

\begin{table}[ht]
    \centering
    \subfloat{
    \begin{tabular}{cccccc}
    \toprule
    \textbf{Car Side} & \textbf{Signal (Human)} & \textbf{$\%$ Save Criminal} & \textbf{N} & \textbf{$\%$ Save Homeless}  & \textbf{N}       \\ 
    \midrule 
    Human & Legal & 0.50 & 1679 & 0.79 & 1630\\
    Human & N/A & 0.44 & 3840 & 0.75 & 3829\\
    Human & Illegal & 0.37 & 2597 & 0.61 & 2526\\
    Dog & Legal & 0.56 & 2571 & 0.82 & 2638\\
    Dog & N/A & 0.52 & 3879 & 0.79 & 3955\\
    Dog & Illegal & 0.43 & 1551 & 0.65 & 1600\\
    \bottomrule
    \end{tabular}}
    \qquad
    \subfloat{
    \begin{tabular}{cccccc}
    \toprule
    \textbf{Car Side} & \textbf{Signal (Human)} & \textbf{$\%$ Save Criminal} & \textbf{N} & \textbf{$\%$ Save Old Man}  & \textbf{N}       \\ 
    \midrule 
    Human & Legal & 0.50 & 1679 & 0.80 & 1659\\
    Human & N/A & 0.44 & 3840 & 0.76 & 3833\\
    Human & Illegal & 0.37 & 2597 & 0.66 & 2538\\
    Dog & Legal & 0.56 & 2571 & 0.83 & 2543\\
    Dog & N/A & 0.52 & 3879 & 0.81 & 3825\\
    Dog & Illegal & 0.43 & 1551 & 0.69 & 1621\\
    \bottomrule
    \end{tabular}}
    \qquad
    \subfloat{
    \begin{tabular}{cccccc}
    \toprule
    \textbf{Car Side} & \textbf{Signal (Human)} & \textbf{$\%$ Save Criminal} & \textbf{N} & \textbf{$\%$ Save Man}  & \textbf{N}       \\ 
    \midrule 
    Human & Legal & 0.50 & 1679 & 0.81 & 1642\\
    Human & N/A & 0.44 & 3840 & 0.79 & 3889\\
    Human & Illegal & 0.37 & 2597 & 0.66 & 2598\\
    Dog & Legal & 0.56 & 2571 & 0.85 & 2597\\
    Dog & N/A & 0.52 & 3879 & 0.83 & 3873\\
    Dog & Illegal & 0.43 & 1551 & 0.69 & 1641\\
    \bottomrule
    \end{tabular}}
    \caption{Results from the Moral Machine dataset corresponding to the scenarios in Experiment 1 / Supplementary Table S9.}
    \label{tab:exp1}
\end{table}

\begin{table}[ht]
    \centering
    \subfloat[][Male]{
    \begin{tabular}{cccccc}
    \toprule
    \textbf{Car Side} & \textbf{Signal (Young)} & \textbf{$\%$ Save (with Adult)} & \textbf{$\%$ Save (without Adult)}    & \textbf{$p$-value} & \textbf{Prediction}       \\ 
    \midrule 
    Young & Legal & 0.83 & 0.77 & $p = .013$ & Null\\
    Young & N/A & 0.77 & 0.75 & $p = .600$ & Null\\
    Young & Illegal & 0.71 & 0.64 & $p = .024$ & Significant\\
    Old & Legal & 0.90 & 0.91 & $p = .827$ & Null\\
    Old & N/A & 0.92 & 0.93 & $p = .538$ & Null\\
    Old & Illegal & 0.81 & 0.75 & $p = .030$ & Significant\\
    \bottomrule
    \end{tabular}}
    \qquad
    \subfloat[][Female]{
    \begin{tabular}{cccccc}
    \toprule
    \textbf{Car Side} & \textbf{Signal (Young)} & \textbf{$\%$ Save (with Adult)} & \textbf{$\%$ Save (without Adult)}    & \textbf{$p$-value} & \textbf{Prediction}       \\ 
    \midrule 
    Young & Legal & 0.84 & 0.80 & $p = .094$ & Null\\
    Young & N/A & 0.78 & 0.76 & $p = .403$ & Null\\
    Young & Illegal & 0.70 & 0.65 & $p = .152$ & Significant\\
    Old & Legal & 0.92 & 0.91 & $p = .562$ & Null\\
    Old & N/A & 0.92 & 0.89 & $p = .061$ & Null\\
    Old & Illegal & 0.81 & 0.73 & $p = .001$ & Significant\\
    \bottomrule
    \end{tabular}}
    \caption{Results from Experiment 2 comparing the percentage of participants that save the young side with an adult versus the percentage of participants that save the young side without the adult. We used a $\chi^2$ analysis between the proportions, where $N = 489$ and $df = 1$.}
    \label{tab:exp2}
\end{table}

\begin{table}[ht]
    \centering
    \subfloat[][Male]{
    \begin{tabular}{cccccc}
    \toprule
    \textbf{Car Side} & \textbf{Signal (Young)} & \textbf{$\%$ Save (with Adult)} & \textbf{N} & \textbf{$\%$ Save (without Adult)}    & \textbf{N}       \\ 
    \midrule 
    Young & Legal & 0.88 & 3516 & 0.88 & 7124\\
    Young & N/A & 0.80 & 8540 & 0.83 & 17289\\
    Young & Illegal & 0.52 & 5745 & 0.64 & 11578\\
    Old & Legal & 0.93 & 5584 & 0.93 & 11428\\
    Old & N/A & 0.92 & 8611 & 0.93 & 17411\\
    Old & Illegal & 0.60 & 3487 & 0.72 & 7299\\
    \bottomrule
    \end{tabular}}
    \qquad
    \subfloat[][Female]{
    \begin{tabular}{cccccc}
    \toprule
    \textbf{Car Side} & \textbf{Signal (Young)} & \textbf{$\%$ Save (with Adult)} & \textbf{N} & \textbf{$\%$ Save (without Adult)}    & \textbf{N}       \\ 
    \midrule 
    Young & Legal & 0.87 & 3589 & 0.89 & 7330\\
    Young & N/A & 0.81 & 8561 & 0.84 & 17193\\
    Young & Illegal & 0.53 & 5743 & 0.66 & 11554\\
    Old & Legal & 0.93 & 5680 & 0.94 & 11399\\
    Old & N/A & 0.92 & 8654 & 0.93 & 17306\\
    Old & Illegal & 0.61 & 3480 & 0.72 & 7166\\
    \bottomrule
    \end{tabular}}
    \caption{Results from the Moral Machine dataset corresponding to the scenarios in Experiment 2 / Supplementary Table S11.}
    \label{tab:exp2}
\end{table}

\begin{table}[ht]
    \centering
    \subfloat[Male-Female Dilemmas]{
    \begin{tabular}{cccccc}
    \toprule
    \textbf{Car Side} & \textbf{Age} & \textbf{$\%$ Save (No Signal)} & \textbf{$\%$ Save (Mean)}    & \textbf{$p$-value} & \textbf{Prediction}       \\ 
    \midrule 
    Male & Adult & 0.14 & 0.25 & $p < .001$ & Significant\\
    Female & Adult & 0.55 & 0.50 & $p = .224$ & Null\\
    Male & Old & 0.19 & 0.27 & $p = .012$ & Significant\\
    Female & Old & 0.55 & 0.50 & $p = .196$ & Null\\
    \bottomrule
    \end{tabular}}
    \qquad
    \subfloat[Fat-Fit Dilemmas]{
    \begin{tabular}{cccccc}
    \toprule
    \textbf{Car Side} & \textbf{Sex} & \textbf{$\%$ Save (No Signal)} & \textbf{$\%$ Save (Mean)}    & \textbf{$p$-value} & \textbf{Prediction}       \\ 
    \midrule 
    Fat & Male & 0.15 & 0.27 & $p < .001$ & Significant\\
    Fit & Male & 0.44 & 0.46 & $p = .609$ & Null\\
    Fat & Female & 0.16 & 0.26 & $p = .003$ & Significant\\
    Fit & Female & 0.47 & 0.42 & $p = .253$ & Null\\
    \bottomrule
    \end{tabular}}
    \caption{Results from Experiment 3 comparing the percentage of participants that save the higher-valued individual in the no crossing signal condition versus the mean of the percentages of participants saving the higher-valued individual in the other two crossing signal conditions. We used a $\chi^2$ analysis between the proportions, in which $N = 326$ and $df = 1$.}
    \label{tab:exp3}
\end{table}

\begin{table}[ht]
    \centering
    \subfloat[Male-Female Dilemmas]{
    \begin{tabular}{ccccc}
    \toprule
    \textbf{Car Side} & \textbf{Age} & \textbf{$\%$ Save (No Signal)} & \textbf{N} & \textbf{$\%$ Save (Mean)} \\ 
    \midrule 
    Male & Adult & 0.19 & 19675 & 0.39\\
    Female & Adult & 0.51 & 19798 & 0.50\\
    Male & Old & 0.24 & 19871 & 0.40\\
    Female & Old & 0.57 & 19937 & 0.53\\
    \bottomrule
    \end{tabular}}
    \qquad
    \subfloat[Fat-Fit Dilemmas]{
    \begin{tabular}{ccccc}
    \toprule
    \textbf{Car Side} & \textbf{Age} & \textbf{$\%$ Save (No Signal)} & \textbf{N} & \textbf{$\%$ Save (Mean)} \\ 
    \midrule 
    Fat & Male & 0.18 & 17222 & 0.37\\
    Fit & Male & 0.45 & 17444 & 0.47\\
    Fat & Female & 0.20 & 17357 & 0.38\\
    Fit & Female & 0.46 & 17347 & 0.47\\
    \bottomrule
    \end{tabular}}
    \caption{Results from the Moral Machine dataset corresponding to the scenarios in Experiment 3 / Supplementary Table S13.}
    \label{tab:exp3}
\end{table}

\begin{table}[!htb]
    \centering
    \caption{Iterations of Bayesian Feature Selection}
    \label{table:bayesmetrics}
    \begin{tabular}{ccccc}
    \toprule
    \bf{Iteration No.} & \bf{No. of Features} & \bf{Accuracy} & \bf{AUC} & \bf{AIC} \\ 
    \midrule
    1 & 1562 & 0.757 & 0.816 & 1.058 \\
    2 & 1251 & 0.757 & 0.816 & 1.053 \\
    3 & 678 & 0.758 & 0.818 & 1.046 \\
    4 & 519 & 0.758 & 0.819 & 1.041 \\
    5 & 372 & 0.758 & 0.819 & 1.041 \\
    6 & 289 & 0.759 & 0.819 & 1.041 \\
    7 & 250 & 0.759 & 0.819 & 1.041 \\
    8 & 235 & 0.759 & 0.819 & 1.041 \\
    9 & 227 & 0.759 & 0.819 & 1.041 \\
    10 & 220 & 0.758 & 0.819 & 1.041 \\
    11 & 212 & 0.758 & 0.819 & 1.041 \\
    12 & 207 & 0.757 & 0.819 & 1.041 \\
    13 & 185 & 0.759 & 0.819 & 1.041 \\
    14 & 181 & 0.758 & 0.819 & 1.040 \\

    \bottomrule
    \end{tabular}
\end{table}

\clearpage

\begin{longtable}{lcc}
\toprule
\textbf{Feature}  & \textbf{Mean} &  \textbf{Standard Deviation} \\
\midrule
Man & 0.692 & 0.008 \\
Woman & 0.865 & 0.005 \\
Pregnant & 0.977 & 0.011 \\
Stroller & 1.034 & 0.024 \\
Old Man & 0.263 & 0.005 \\
Old Woman & 0.365 & 0.005 \\
Boy & 1.130 & 0.007 \\
Girl & 1.291 & 0.004 \\
Homeless & 0.404 & 0.005 \\
Large Woman & 0.677 & 0.003 \\
Large Man & 0.428 & 0.006 \\
Male Executive & 0.691 & 0.010 \\
Female Executive & 0.801 & 0.006 \\
Female Athlete & 0.961 & 0.005 \\
Male Athlete & 0.768 & 0.005 \\
Female Doctor & 0.860 & 0.006 \\
Male Doctor & 0.821 & 0.008 \\
Dog & 0.152 & 0.004 \\
Crossing Signal & 0.950 & 0.004 \\
Car Side & -0.274 & 0.005 \\
Woman * Female Doctor & -0.132 & 0.009 \\
Old Man * Criminal & 0.085 & 0.015 \\
Old Man * Dog & -0.086 & 0.008 \\
Old Woman * Boy & 0.040 & 0.005 \\
Old Woman * Girl & 0.020 & 0.005 \\
Boy * Female Doctor & 0.084 & 0.009 \\
Girl * Female Doctor & -0.057 & 0.008 \\
Girl * Crossing Signal & -0.035 & 0.008 \\
Large Woman * Male Athlete & 0.061 & 0.009 \\
Large Man * Dog & -0.057 & 0.008 \\
Criminal * Cat & -0.052 & 0.008 \\
Female Athlete * Crossing Signal & -0.049 & 0.006 \\
Man * Woman * Old Man & 0.025 & 0.004 \\
Man * Woman * Old Woman & -0.014 & 0.005 \\
Man * Woman * Large Man & -0.021 & 0.005 \\
Man * Woman * Cat & 0.176 & 0.022 \\
Man * Stroller * Female Executive & -0.192 & 0.021 \\
Man * Old Woman * Male Athlete & -0.118 & 0.010 \\
Man * Boy * Crossing Signal & 0.055 & 0.012 \\
Man * Girl * Male Executive & -0.146 & 0.023 \\
Man * Girl * Female Executive & -0.141 & 0.013 \\
Man * Girl * Female Athlete & -0.206 & 0.017 \\
Man * Large Man * Male Executive & -0.213 & 0.017 \\
Man * Large Man * Female Doctor & -0.196 & 0.020 \\
Man * Female Executive * Female Athlete & -0.121 & 0.023 \\
Man * Male Athlete * Crossing Signal & -0.047 & 0.007 \\
Man * Male Doctor * Dog & 0.055 & 0.016 \\
Man * Dog * Cat & -0.047 & 0.009 \\
Woman * Pregnant * Boy & -0.181 & 0.026 \\
Woman * Pregnant * Criminal & -0.220 & 0.056 \\
Woman * Pregnant * Female Athlete & -0.239 & 0.018 \\
Woman * Pregnant * Male Athlete & -0.293 & 0.029 \\
Woman * Pregnant * Male Doctor & -0.467 & 0.022 \\
Woman * Stroller * Cat & 0.159 & 0.020 \\
Woman * Old Man * Female Athlete & -0.210 & 0.015 \\
Woman * Old Man * Male Athlete & -0.148 & 0.012 \\
Woman * Old Woman * Dog & 0.216 & 0.019 \\
Woman * Boy * Male Athlete & -0.269 & 0.028 \\
Woman * Boy * Female Doctor & -0.181 & 0.011 \\
Woman * Girl * Female Doctor & -0.353 & 0.009 \\
Woman * Girl * Male Doctor & -0.179 & 0.015 \\
Woman * Homeless * Cat & 0.134 & 0.012 \\
Woman * Large Woman * Female Executive & -0.177 & 0.015 \\
Woman * Large Woman * Cat & 0.054 & 0.009 \\
Woman * Large Man * Male Athlete & -0.034 & 0.004 \\
Woman * Large Man * Female Doctor & -0.119 & 0.018 \\
Woman * Criminal * Female Doctor & -0.275 & 0.034 \\
Woman * Male Executive * Female Doctor & -0.189 & 0.016 \\
Woman * Male Executive * Male Doctor & -0.321 & 0.023 \\
Woman * Female Executive * Female Athlete & -0.254 & 0.015 \\
Woman * Female Executive * Female Doctor & -0.182 & 0.026 \\
Woman * Female Executive * Male Doctor & -0.271 & 0.015 \\
Woman * Female Athlete * Female Doctor & -0.292 & 0.019 \\
Woman * Female Athlete * Male Doctor & -0.146 & 0.017 \\
Woman * Male Athlete * Female Doctor & -0.057 & 0.026 \\
Pregnant * Stroller * Male Doctor & -0.265 & 0.051 \\
Pregnant * Old Man * Boy & 0.077 & 0.017 \\
Pregnant * Old Man * Male Athlete & 0.072 & 0.019 \\
Pregnant * Old Man * Cat & 0.078 & 0.024 \\
Pregnant * Old Woman * Criminal & -0.342 & 0.028 \\
Pregnant * Old Woman * Male Executive & -0.323 & 0.029 \\
Pregnant * Boy * Girl & -0.251 & 0.028 \\
Pregnant * Girl * Male Doctor & -0.316 & 0.032 \\
Pregnant * Homeless * Male Doctor & -0.183 & 0.049 \\
Pregnant * Female Doctor * Male Doctor & -0.123 & 0.051 \\
Pregnant * Female Doctor * Crossing Signal & 0.170 & 0.036 \\
Pregnant * Dog * Cat & -0.117 & 0.006 \\
Stroller * Old Woman * Boy & -0.196 & 0.045 \\
Stroller * Old Woman * Girl & -0.144 & 0.035 \\
Stroller * Boy * Crossing Signal & 0.152 & 0.016 \\
Stroller * Girl * Male Executive & -0.188 & 0.025 \\
Stroller * Girl * Crossing Signal & 0.289 & 0.039 \\
Stroller * Homeless * Crossing Signal & -0.131 & 0.026 \\
Stroller * Large Woman * Male Executive & -0.112 & 0.054 \\
Stroller * Large Woman * Male Doctor & -0.135 & 0.035 \\
Stroller * Female Executive * Male Athlete & -0.321 & 0.028 \\
Stroller * Female Athlete * Crossing Signal & 0.176 & 0.018 \\
Stroller * Female Doctor * Male Doctor & -0.335 & 0.026 \\
Stroller * Dog * Cat & -0.115 & 0.007 \\
Stroller * Dog * Crossing Signal & 0.109 & 0.009 \\
Old Man * Old Woman * Boy & -0.082 & 0.005 \\
Old Man * Old Woman * Girl & -0.136 & 0.006 \\
Old Man * Old Woman * Female Doctor & 0.074 & 0.008 \\
Old Man * Old Woman * Crossing Signal & -0.047 & 0.006 \\
Old Man * Boy * Crossing Signal & -0.044 & 0.008 \\
Old Man * Girl * Female Doctor & -0.114 & 0.017 \\
Old Man * Girl * Crossing Signal & -0.114 & 0.008 \\
Old Man * Male Executive * Female Executive & -0.101 & 0.007 \\
Old Man * Male Athlete * Female Doctor & -0.153 & 0.018 \\
Old Man * Female Doctor * Crossing Signal & -0.090 & 0.014 \\
Old Woman * Boy * Girl & 0.060 & 0.005 \\
Old Woman * Boy * Large Man & 0.070 & 0.021 \\
Old Woman * Girl * Male Executive & -0.053 & 0.027 \\
Old Woman * Girl * Female Doctor & -0.174 & 0.018 \\
Old Woman * Large Woman * Male Executive & -0.116 & 0.016 \\
Old Woman * Large Woman * Female Executive & -0.223 & 0.015 \\
Old Woman * Large Man * Criminal & 0.119 & 0.019 \\
Old Woman * Large Man * Male Executive & -0.109 & 0.019 \\
Boy * Girl * Female Athlete & 0.041 & 0.008 \\
Boy * Large Woman * Female Athlete & -0.094 & 0.013 \\
Boy * Large Woman * Male Athlete & -0.238 & 0.011 \\
Boy * Large Woman * Crossing Signal & 0.047 & 0.019 \\
Boy * Large Man * Crossing Signal & 0.128 & 0.012 \\
Boy * Male Executive * Crossing Signal & 0.130 & 0.013 \\
Boy * Female Executive * Male Doctor & 0.068 & 0.009 \\
Boy * Female Executive * Dog & 0.127 & 0.017 \\
Boy * Female Executive * Crossing Signal & 0.120 & 0.012 \\
Boy * Female Athlete * Crossing Signal & 0.178 & 0.015 \\
Boy * Male Athlete * Crossing Signal & 0.191 & 0.017 \\
Boy * Female Doctor * Crossing Signal & 0.255 & 0.014 \\
Boy * Male Doctor * Crossing Signal & 0.152 & 0.016 \\
Boy * Dog * Cat & -0.126 & 0.005 \\
Boy * Dog * Crossing Signal & 0.169 & 0.014 \\
Girl * Homeless * Female Doctor & -0.256 & 0.034 \\
Girl * Large Woman * Large Man & 0.052 & 0.010 \\
Girl * Large Woman * Male Athlete & -0.174 & 0.014 \\
Girl * Large Man * Cat & 0.157 & 0.012 \\
Girl * Criminal * Female Doctor & -0.365 & 0.016 \\
Girl * Male Executive * Female Executive & 0.067 & 0.009 \\
Girl * Male Executive * Cat & 0.073 & 0.023 \\
Girl * Male Executive * Crossing Signal & 0.203 & 0.020 \\
Girl * Female Executive * Dog & 0.211 & 0.014 \\
Girl * Female Executive * Crossing Signal & 0.231 & 0.019 \\
Girl * Female Athlete * Crossing Signal & 0.262 & 0.014 \\
Girl * Male Athlete * Cat & 0.072 & 0.014 \\
Girl * Male Athlete * Crossing Signal & 0.276 & 0.010 \\
Girl * Female Doctor * Crossing Signal & 0.183 & 0.010 \\
Girl * Male Doctor * Crossing Signal & 0.088 & 0.020 \\
Girl * Dog * Cat & -0.154 & 0.005 \\
Girl * Cat * Crossing Signal & 0.188 & 0.014 \\
Homeless * Male Executive * Crossing Signal & 0.056 & 0.015 \\
Homeless * Female Executive * Cat & 0.116 & 0.024 \\
Homeless * Cat * Crossing Signal & 0.115 & 0.014 \\
Large Woman * Large Man * Female Athlete & -0.030 & 0.005 \\
Large Woman * Large Man * Male Athlete & -0.047 & 0.005 \\
Large Woman * Female Athlete * Male Athlete & 0.038 & 0.005 \\
Large Woman * Male Doctor * Dog & 0.070 & 0.027 \\
Large Woman * Dog * Cat & -0.056 & 0.004 \\
Large Man * Male Executive * Dog & 0.095 & 0.010 \\
Large Man * Female Doctor * Cat & 0.085 & 0.013 \\
Male Executive * Female Executive * Dog & 0.105 & 0.014 \\
Male Executive * Female Athlete * Crossing Signal & 0.162 & 0.010 \\
Male Executive * Male Athlete * Cat & 0.148 & 0.014 \\
Male Executive * Female Doctor * Cat & 0.154 & 0.026 \\
Male Executive * Dog * Crossing Signal & 0.062 & 0.021 \\
Female Executive * Female Athlete * Crossing Signal & 0.255 & 0.015 \\
Female Executive * Male Athlete * Male Doctor & -0.211 & 0.012 \\
Female Executive * Female Doctor * Cat & 0.151 & 0.020 \\
Female Executive * Female Doctor * Crossing Signal & 0.154 & 0.015 \\
Female Executive * Male Doctor * Dog & 0.096 & 0.018 \\
Female Executive * Male Doctor * Crossing Signal & 0.100 & 0.010 \\
Female Executive * Dog * Crossing Signal & 0.155 & 0.007 \\
Female Athlete * Male Doctor * Cat & 0.095 & 0.013 \\
Female Athlete * Dog * Cat & -0.108 & 0.007 \\
Female Athlete * Cat * Crossing Signal & 0.165 & 0.009 \\
Male Athlete * Female Doctor * Dog & 0.175 & 0.016 \\
Male Athlete * Female Doctor * Cat & 0.123 & 0.018 \\
Male Athlete * Male Doctor * Cat & 0.222 & 0.010 \\
Male Athlete * Dog * Crossing Signal & 0.120 & 0.013 \\
Female Doctor * Male Doctor * Cat & 0.230 & 0.013 \\
Female Doctor * Cat * Crossing Signal & 0.153 & 0.013 \\
Male Doctor * Cat * Crossing Signal & 0.094 & 0.013 \\
\caption{Mean and Standard Deviation of Posterior Weights for Bayesian Variable Selection} % needs to go inside longtable 
\end{longtable}

\clearpage

\end{document}